\pdfoutput=1

\newcommand\Msun{\hbox{M$_\odot$}}
\newcommand\Zsun{\hbox{Z$_\odot$}}
\newcommand\kms{\hbox{$\,$km$\,$s$^{-1}$}}

\newcommand\one{\,{\sc i}}
\newcommand\two{\,{\sc ii}}
\newcommand\three{\,{\sc iii}}

\newcommand\tmult{\multicolumn{2}{c}}
\newcommand\hst{\textit{HST}}
\newcommand\chan{\textit{Chandra}}
\newcommand\spit{\textit{Spitzer}}
\newcommand\twomass{\textit{2MASS}}
\newcommand\ie{\textit{i.\,e.}}
\newcommand\eg{e.\,g.}
\newcommand\cf{\textit{cf.}}
\newcommand\etal{et~al.}

\newcommand\bb{$B_{435}$}
\newcommand\vb{$V_{606}$}
\newcommand\ib{$I_{814}$}

\newcommand\ratios{$\frac{\log(M_{\textup{\scriptsize H\one}})}{\log(M_{\textup{\scriptsize dyn}})}$}
\newcommand\airac{$\alpha_\textup{\scriptsize IRAC}$}

\newcommand\gala{\textit A}
\newcommand\galb{\textit B}
\newcommand\galc{\textit C}
\newcommand\gald{\textit D}
\newcommand\galx{\textit E}
\newcommand\galacd{\textit{A/C/D}}
\newcommand\mstar{$M_*$}
\newcommand\mhi{$M_\textup{\scriptsize H\one}$}
\newcommand\nhi{$N_\textup{\scriptsize H\one}$}
\newcommand\mhtwo{$M_{\textup{\scriptsize H}_2}$}
\newcommand\xco{$X_{\textup{\scriptsize CO}}$}
\newcommand\tquench{$\tau_\textup{\scriptsize depletion}$}
\newcommand\ha{H$\alpha$}
\newcommand\hb{H$\beta$}

\newcommand\ned{\textit{NED}}
\newcommand{\farcs}{\mbox{\ensuremath{.\!\!^{\prime\prime}}}}

\documentclass{emulateapj}
\usepackage{graphicx,epstopdf,rotating,natbib}
\shorttitle{Hickson Compact Group 7}
\shortauthors{Konstantopoulos et al.}

\begin{document}

\title{Galaxy evolution in a complex environment: a multi-wavelength study of HCG~7\footnote{Based on observations made with the NASA/ESA Hubble Space Telescope}}

\author{I.~S.~Konstantopoulos\altaffilmark{1},
S.~C. Gallagher\altaffilmark{2},
K.~Fedotov\altaffilmark{2},
P.~R.~Durrell\altaffilmark{3},
A.~Heiderman\altaffilmark{4},
D.~M.~Elmegreen\altaffilmark{5},
J.~C. Charlton\altaffilmark{1},
J.~E.~Hibbard\altaffilmark{10},
P.~Tzanavaris\altaffilmark{6,7},
R.~Chandar\altaffilmark{8},
K.~E.~Johnson\altaffilmark{9,10},
A.~Maybhate\altaffilmark{11},
A.~E.~Zabludoff\altaffilmark{12},
C.~Gronwall\altaffilmark{1},
D.~Szathmary\altaffilmark{2},
A.~E.~Hornschemeier\altaffilmark{6},
J.~English\altaffilmark{13},
B.~Whitmore\altaffilmark{11},
C Mendes de Oliveira \altaffilmark{14},
J S Mulchaey \altaffilmark{15}
}

 \altaffiltext{1}{Department of Astronomy and Astrophysics, The Pennsylvania State University, University Park, PA 16802; iraklis@psu.edu}
 \altaffiltext{2}{Department of Physics \& Astronomy, The University of Western Ontario, London, ON, N6A 3K7, Canada} 
 \altaffiltext{3}{Department of Physics \& Astronomy, Youngstown State University, Youngstown, OH 44555}
 \altaffiltext{4}{Department of Astronomy, University of Texas at Austin, Austin, TX 78712-0259}
 \altaffiltext{5}{Department of Physics \& Astronomy, Vassar College, Poughkeepsie, NY 12604}
 \altaffiltext{6}{Laboratory for X-ray Astrophysics, NASA's Goddard Space Flight Center, Greenbelt, MD 20771}
 \altaffiltext{7}{Department of Physics and Astronomy, The Johns Hopkins University, Baltimore, MD 21218}
 \altaffiltext{8}{University of Toledo, Toledo, OH}
 \altaffiltext{9}{University of Virginia, Charlottesville, VA}
 \altaffiltext{10}{National Radio Astronomy Observatory, Charlottesville, VA}
 \altaffiltext{11}{Space Telescope Science Institute, Baltimore, MD}
 \altaffiltext{12}{Steward Observatory, University of Arizona, Tucson, AZ 85721}
 \altaffiltext{13}{University of Manitoba, Winnipeg, MN, Canada}
 \altaffiltext{14}{Universidade de S\~ao Paulo, IAG, S\~ao Paulo, SP 05508-900, Brazil}
 \altaffiltext{15}{Carnegie Observatories, Pasadena, CA  91101}

\begin{abstract}
The environment where galaxies are found heavily influences their evolution. Close groupings, like the ones in the cores of galaxy clusters or compact groups, evolve in ways far more dramatic than their isolated counterparts. We have conducted a multi-wavelength study of Hickson~Compact~Group~7, consisting of four giant galaxies: three spirals and one lenticular. We use \textit{Hubble Space Telescope} (\hst) imaging to identify and characterize the young and old star cluster populations. We find young massive clusters (YMC) mostly in the three spirals, while the lenticular features a large, unimodal population of globular clusters (GC) but no detectable clusters with ages less than a few Gyr. The spatial and approximate age distributions of the \mbox{$\sim300$ YMCs} and \mbox{$\sim150$ GCs} thus hint at a regular star formation history in the group over a Hubble time. While {at first glance the \hst\ data show the galaxies as undisturbed}, our deep ground-based, wide-field imaging {that extends the \hst\ coverage} reveals faint signatures of stellar material in the intra-group medium. {We do not, however, detect the intra-group medium in H\one\ or \chan\ X-ray observations, signatures that would be expected to arise from major mergers}. Despite this fact, we find that the H\one\ gas content of the individual galaxies and the group as a whole are a third of the expected abundance. The appearance of quiescence is {challenged} by spectroscopy that reveals {an intense ionization continuum in one galaxy nucleus, and post-burst characteristics in another}. Our spectroscopic survey of dwarf galaxy members yields a single dwarf elliptical galaxy in an apparent stellar tidal feature. Based on all this information, we {suggest an evolutionary scenario for HCG~7}, whereby {the galaxies convert most} of their available gas {into stars} without the influence of major mergers and ultimately result in a dry merger. As the conditions governing compact groups are reminiscent of galaxies at intermediate redshift, we propose that HCGs are appropriate for studying galaxy evolution at $z\sim1-2$.
\end{abstract}


\section{Introduction}
The high luminosities and impressive, disturbed morphologies of local
merging and interacting galaxies testify to the ability of these
processes to drive rapid galaxy evolution. 
Such events often result in increased star formation activity\footnote{
{This might not always be the case, however, as challenged by
\citet{bergvall03}, who presented mergers as scaled-up versions of 
isolated spirals, without a necessary SFR enhancement. 
}}, 
as testified by the enhanced infrared (IR) luminosities \citep{sanders96}, 
and the redistribution of the gaseous and stellar components of the individual galaxies
\citep{holtzman92,hibbard96,schweizer98,zhang99,whitmore99,gelys07a,isk09a}. 
Galaxies, however, are also found in isolation. In the absence of external
influences, including tidal forces from massive galaxies and
interactions with intracluster or intragroup media (IGM), these true `island
universes' can evolve through internal processes such as disk
instabilities that boost star formation and funnel gas to smaller
radii \citep[e.g.,][and references therein]{kk04}. In cases where internal dynamics dominate
external processes over cosmological timescales, we speak of secular
(\ie~continuous and cumulative) evolution.  The relative importance of
these two processes is under debate. Indeed, the latter may have been
discounted during the past two decades of \hst-driven studies of
intensely star-forming environments. Under both scenarios, dynamical processes in concert with the
conversion of gas into stars can lead to the morphological evolution
of disk-dominated galaxies from gas-rich and star-forming to quiescent
and featureless with substantial bulge growth.  Interestingly, the 
mechanisms that drive this transformation are far from
certain. Do mergers and interactions represent the main path from star-forming
spirals to quiescent ellipticals \citep[\eg][]{genzel01}, 
or does secular evolution
represent an equally relevant (at $z\sim0$), albeit less eventful
path?

Most galaxies are not found at the extremes of true isolation in the
field or conversely within galaxy cluster cores, but in the intermediate density
environments of groups \citep{eke04}.  In such systems, both types of evolutionary mechanisms
are likely to be operative; many galaxies will have undergone
interactions and mergers, but the dynamical times are long enough that
not all galaxies will be affected.  Compact galaxy groups, with
several galaxies within a few galaxy radii, offer excellent
arenas for investigating morphological evolution triggered by
interactions at high number density; their low velocity dispersions, 
$\sigma\sim10^2\kms$ \citep{tago08}, with respect to cluster {cores} 
\citep[$\sim10^3\kms$, \eg][]{binggeli87,the86} extend the timescales of
gravitational encounters.  Furthermore, the intragroup medium in all
but the most massive groups is insufficient for ram-pressure
stripping, which simplifies the investigation of the agents of
morphological change in comparison with clusters \citep{rasmussen08}.

Hickson Compact Groups \citep[HCG;][]{hickson82,hickson92} provide
well-studied examples of this fruitful environment.  Initially
selected based on high number density and isolation (to exclude
clusters), they are found to be generally H\one\ deficient
when compared to field galaxies of matched morphologies
\citep{williams87,verdes01}, implying that acclerated gas consumption
is already underway.
HCGs as a population are diverse, and range from conglomerations of
small, gas-rich galaxies apparently coming together for the first time
such as HCG~31 \citep[e.g.,][]{gallagher10,rubin91} to
massive, highly evolved and X-ray bright groups such as HCG~42 that
are dominated by quiescent ellipticals \citep[e.g.,][]{zab98}.
Another interesting trait exhibited by HCGs is a gap in the
mid-IR color distribution of member galaxies
\cite[][hereafter J07 and W09]{johnson07,walker09} that is not
present in test samples of field galaxies. This can be interpreted as
evidence of rapid evolution from gas-rich and star-forming to gas-poor
and quiescent, possibly accompanied by the build-up
of a bulge.  At the same time, the high fraction ($\sim43\%$) of
optically evident morphological disturbances in compact group galaxies
\citep{mendes94} does not map onto correspondingly high star formation
rates from infrared surveys \citep{moles94,vm98}.  In some sense, the
dynamical state (i.e., the strength of morphological interaction
indicators) of a compact group is not a good predictor of its ongoing
star formation activity \citep{iglesias}.  Therefore, the history of
direct collisions does not provide the whole story.

\subsection{An evolutionary sequence for compact galaxy groups}\label{sec:seq}
\begin{figure*}
\begin{center}

	\includegraphics[width=\textwidth, angle=0]{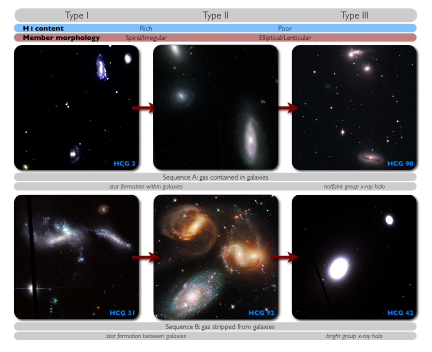}
	
	\caption{The proposed double-branched evolutionary sequence for compact galaxy groups. The upper sequence shows the evolution of groups in the absence of gas in the IGM. Under this scenario, star formation proceeds uninterrupted in the individual galaxies and the H\one\ gas is consumed before major interactions can take place within the group. Thus, gas does not play a significant role in the ensuing `dry merger'. The lower sequence shows a {better studied} situation, where a number of strong interactions take place at early times in the course of  the group's evolution. In that way, the IGM is enriched with gas that is used partly in forming stars, while another part {contributes to the X-ray IGM}. Our measurements of the way in which gas is  processed  in HCGs may be related to the $L_X/L_B$ observed in the current-era universe. 
	}\label{fig:sequence}
	
\end{center}
\end{figure*}

Our previous work has dealt with compact groups both individually
\citep{gallagher10} and in the ensemble
\citep[J07; W09;][]{gallagher08,tzanavaris10}. Our sample is composed of the 12
groups in the Hickson catalog within 4500~km~s$^{-1}$ with at least 3 accordant members that
are compact enough to fit within a few arcminutes (for efficient
observing).  To investigate the evolutionary stage of individual
groups as a whole, J07 classified them as early, intermediate or late
following the qualitative scheme of \citet{verdes01} according to
their neutral gas content (from rich to poor) and morphological state
(from spiral-dominated to E/SO-dominated) of the galaxies.  In tandem,
groups were divided quantitatively into three categories according to
the total gas available to the system divided by the dynamical
mass, as presented in Fig.~\ref{fig:sequence}. This ratio of gas-to-dynamical mass may serve as a
proxy of evolutionary stage, though the mapping is not one-to-one. 
J07 classified three categories: \\
(I) relatively H\one-rich, \ratios~$\geq0.9$;\\ 
(II) intermediate H\one, $0.9>$~\ratios$~\geq0.8$;\\
(III) relatively H\one-poor, \ratios~$<0.8$.

\subsection{Lenticular galaxies in the group environment}
From their recent comparison of $z\sim0.4$ field and group galaxies,
\citet{wilman} conclude that the high S0 fraction in galaxy clusters
is driven by infalling groups, but the lenticulars are forged through
preprocessing in the group environment.  Some intriguing results from
local compact group studies suggest that this environment may offer
fertile ground for exploring this issue.  For example, a kinematic
study of HCG~90 by \citet{plana98} revealed that two member galaxies,
HCG~90b and d, morphologically classified as ellipticals, have
rotating stellar disks and are more accurately classified as S0s.  This group
is also notable for its large fraction ($\sim38-48\%$) of
intracluster light compared to total group light and very faint
X-ray-emitting intragroup medium \citep{white}.  Similarly, Seyfert's
Sextet (HCG~79), obviously undergoing intense and complex interactions
from its highly distorted member galaxies, revealed a surprising lack
of young star clusters \citep{palma} compared to the similarly
disrupted Stephan's Quintet \citep[HCG~92;][]{gall01}.  In this paper,
we turn to HCG~7 to flesh out the intermediate stages of one type of compact
group evolution that plausibly results in systems such as HCG~79 and
90.

\subsection{HCG~7}
\begin{figure*}
\begin{center}

	\includegraphics[width=\textwidth, angle=0]{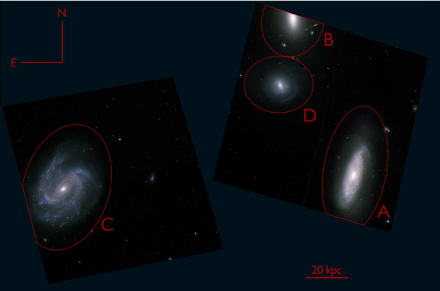}\\
	\includegraphics[width=\textwidth, angle=0]{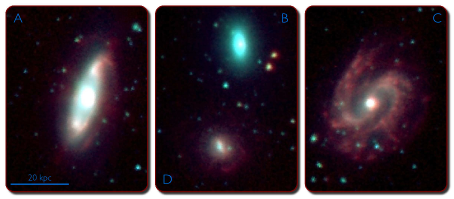}
	
	\caption{
	\small{\bf Top: }\hst\ \textit{BVI} color composite image of the four galaxies in two pointings (see Table~\ref{tab1} for basic information). The image covers all four group member galaxies. \gala, \galc\ and \gald\ (NGC~192, 201 and 197 respectively) are barred spiral galaxies; \gala\ is highly inclined, while \gald\ projects a rather limited extent and very faint spiral structure. \galb\ (NGC~196) is a barred lenticular galaxy (SB0). The ellipses indicate the regions we have associated to each of the galaxies; any objects found outside those areas are considered to occupy the intra-cluster medium. From the optical image alone, it is obvious that the four galaxies are `strangers', in the sense that they show no record of past interactions. {\bf Bottom: } \spit\ imaging, color-coded with blue, green and red colors corresponding to 3.6, 4.5 and 8$~\mu$m. In that way, photospheric emission from the old stellar population shines blue, while star-forming regions are revealed by the bright red PAH emission. \normalsize
	}
	\label{fig:finder}
	
\end{center}
\end{figure*}

Group 7 in the Hickson catalog consists of four galaxies: three
barred spirals, NGC~192, NGC~201 and NGC~197 (members A, C and D), and
one barred lenticular, NGC~196 (B). {We will refer to the the galaxies by their alphabetical classifiers, 
\gala, \galb, \galc\ and \gald\ throughout this paper}. We present
new \hst\ imaging of the group in two pointings in
Fig.~\ref{fig:finder}. \gala\ is highly inclined, while \gald\
projects a rather limited extent and faint, asymmetric spiral structure. 
The imposing \galc\
displays a clear, face-on view of its two dominant spiral arms, one of
which may have a composite structure. Positional, morphological,
photometric and redshift information has been included in
Table~\ref{tab1}.

\begin{table*}[htbp]
\begin{center}
\caption{Basic information on HCG~7 member galaxies\tablenotemark{a}}\label{tab1}
\begin{tabular}{lccccccc} 
\tableline
\tableline
Identifier	& Coordinates	& \tmult{Type}  & $m$\tablenotemark{a} & $v_R$\tablenotemark{b}\\
		& (J2000)	& H89	\tablenotemark{c}	& {RC3} & (mag) & (\kms)\\
\tableline
	A: NGC~192	& 00~39~13.4 +00~51~51	& SB 	& (RÕ)SB(r)a & 13.4 ($V$) & 4160\\
	B: NGC~196	& 00~39~17.8 +00~54~46	& SB0	& SB0pec	& 13.8 ($V$) & 4135\\
	C: NGC~201	& 00~39~34.8 +00~51~36	& SBc	& SAB(r)c	& 13.8 ($g$) & 4356\\
	D: NGC~197	& 00~39~18.8 +00~53~31	& SBc	& SB0pec	& 15.4 ($g$) & 4107\\
	E: $-$			& 00 39 15.5 +00~56~33		& $-$	& dE\tablenotemark{d} & 18.5 & 4118\\
\tableline
\end{tabular}
\tablenotetext{a}{From the Third Reference Catalogue of Bright Galaxies, \citet[][RC3]{rc3}. }
\tablenotetext{b}{Measured in this work, using the RVSAO package in IRAF. }
\tablenotetext{c}{From \citet{hickson89}. }
\tablenotetext{d}{As derived in this work, not from RC3; see \S~\ref{sec:dwarf}.}
\end{center}
\end{table*}

Perhaps most interestingly, all spirals in this group appear to be
undisturbed, thus HCG~7 offers the
opportunity for an excellent case study of galaxy evolution in the
group environment in the absence of obvious strong interactions.
This work forms part of a series of papers that use HCGs as probes of
galaxy evolution in dense environments \citep[J07; W09][]{gallagher08,gallagher10,tzanavaris10}. Specifically, we are applying an extensive,
multi-wavelength dataset of a sample of 12 nearby HCGs to the problem
of characterizing star formation and following gas consumption and
distribution within compact galaxy groups.

Over the following sections, we present an in-depth study of the
group, assess its current state, interpret its past and attempt to
predict its evolution. We will provide the first high resolution study
of the young stellar populations and cluster complexes in HCG~7, and
the first study of globular clusters in a compact group
lenticular. This will then be put in the context of galaxy
evolution by comparison to field galaxies and other compact groups. The
following two sections (Sections~\ref{sec:data},~\ref{sec:moredata}) present the optical and
infrared imaging, radio data, and optical spectroscopy we
have used to dissect this intriguing system. 
Section~\ref{sec:s-pops} includes the analysis of the stellar
content at all levels of the star formation hierarchy, \ie\ young and
old star clusters, cluster forming regions/complexes and dwarf
galaxies. 
The globular cluster population is treated separately in Section~\ref{sec:gcs}. 
An interpretation of the multi-wavelength properties of each group galaxy is 
presented in Section~\ref{sec:seds}. 
We discuss the implications of our findings in
Section~\ref{sec:discussion}, where we also explore the significance
of HCG~7 in the context of compact group evolution. We summarize our
findings in Section~\ref{sec:summary}.
{The analysis presented in this paper assumes the set of cosmological 
parameters of \citet{spergel07}: $H_0=0.70$, $\Omega_m=0.3$, and $\Omega_\lambda=0.7$}.

\section{\hst\ ACS imaging and cluster candidate selection}\label{sec:data}
The basis of our analysis of HCG~7 is the \hst\ multi-band imaging, obtained using the Wide Field Channel (WFC) of the Advanced Camera for Surveys (ACS), in the \textit{F435W, F606W \textup{and} F814W} bands and in two pointings (to cover all group members). We adopt the notation \bb, \vb, \ib\ to relate the observations to the Johnson photometric system, however, the notation does not imply a conversion between the two systems. The observations were executed on 10 and 11 September 2006, as part of program 10787 (PI: Jane Charlton). The {total} exposure times were 1710, 1230 and 1065 seconds in the \textit{BVI} bands respectively. The observations for each filter were taken with three equal exposures, using a three-point dither pattern (sub-pixel dithering). The implementation of \texttt{MultiDrizzle} \citep{multidrizzle} in the \hst/ACS data pipeline provides combined, geometrically corrected, and cosmic-ray cleaned images. For the analysis of point sources, we used the standard \hst\ pipeline products with a nominal pixel scale of $0\farcs05$ per pixel. For analysis of the extended sources, we ran \texttt{MultiDrizzle} with the pixel scale set to $0\farcs03$ per pixel to improve the spatial resolution. 

The images, presented in Fig.~\ref{fig:finder} (top panel), were used in a number of ways: to define the optical extent of galaxy disks and probe star formation in the IGM; to detect stellar aggregates and sub-galactic structures (\ie\ individual star clusters and cluster complexes); to obtain the magnitudes and colors of all these objects, and therefore estimate their ages. This enables the distinction between young and old clusters, \ie\ the bright, blue young massive clusters (YMCs; also referred to as super star clusters, or SSCs) and fainter, red globular clusters (GCs). 

\subsection{Selection and photometry of star cluster candidates}
At the adopted distance to HCG~7 of 56.6~Mpc \citep{hickson92}, {the only stars that are luminous enough to be detected are supergiants}. Instead, star clusters are largely used as alternatives to individual stars in studying extragalactic star formation \citep[\eg][]{gelys07b,isk08,isk09a,isk09b,bastian09antennae}. Star clusters maintain a link to the overall star formation in any system \citep{ladalada03} and can therefore be used to investigate the star formation history \citep[\eg][]{oestlin98,anders04b}. Careful selection is essential, however, as clusters will appear as point sources on these images -- at $56.6$~Mpc, one pixel on the ACS chip corresponds to $\sim11$~pc, \cf\ the average star cluster {half-light} radius of $\sim4$~pc, {as measured in a variety of environments, \eg\ M101 \citep{barmby06}, M51 \citep{remco07}, and a selection of star-forming spirals \citep{larsen04}}. 

	We used the \texttt{DAOphot} package in IRAF\footnote{
	IRAF is distributed by the National Optical Astronomy Observatories,
	which are operated by the Association of Universities for Research
	in Astronomy, Inc., under cooperative agreement with the National
	Science Foundation.} to detect sources on the \hst\ images, following the process presented in 
	\citet{gallagher10}: instead of applying a high detection threshold, we used a median-divided image to perform the initial selection. This prevents the loss of genuine clusters in high or varying backgrounds. We used a square of side 13 pixels to smooth the image and then divide the original by this smoothed image. Thus, the background is effectively lowered to unity (background divided by background) and the unresolved, marginally resolved, and extended sources can be distinguished. This also acts as a first filter against artifacts, as bright sources of limited extent will be effectively smeared out. After completing this transformation, we can use a low detection threshold to detect sources on the median-divided image.

A number of filters were applied to a detected source before it could be considered a star cluster candidate (SCC), which at this point did not differentiate between YMCs and GCs. At this first stage we used the \hst\ weight maps to exclude spurious sources and we also restricted the selection to sources detected in all three filters. 

\begin{figure*}
\begin{center}

	\includegraphics[width=\textwidth]{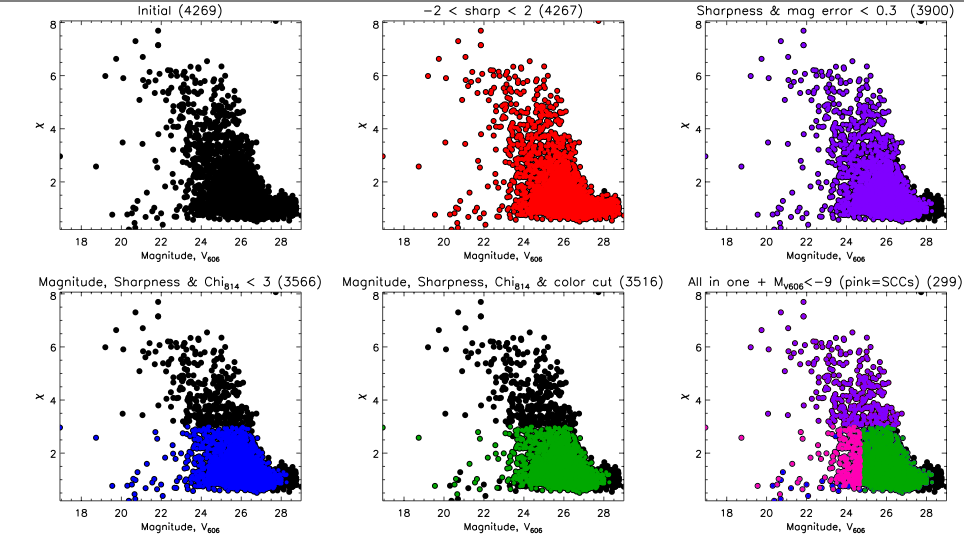}
	
	\caption{Source filtering through the diagnostic measures output by \texttt{DAOphot.allstar}. Before this step we combine the \bb, \vb\ and \ib\ catalogs to reject spurious sources. We then apply cuts in source `sharpness' (effectively a source concentration index), photometric error and according to the allstar $\chi$ diagnostic, before employing colour and magnitude cuts appropriate for star clusters. This stringent set of filters yield a somewhat low number of sources, which, however, make up a sample of \textit{bona-fide} clusters. 
	}\label{fig:filters}
	
\end{center}
\end{figure*}

	{We then proceed to apply a number of filters to eliminate sources of spurious or stellar nature, as well as background galaxies. We demonstrate the successive application of filters in Fig.~\ref{fig:filters} and discuss the selection criteria below. Before applying the filters described in the following text, we cross-correlate the \bb, \vb\ and \ib\ catalogues to remove spurious sources. }

	With star clusters appearing as point sources, point spread function (PSF) fitting needs to be conducted to obtain their photometric properties. A PSF was constructed for each filter using a number of bright, isolated stars with clear, smooth growth curves. We used \texttt{daophot.allstar} to conduct PSF photometry and applied aperture corrections to the photometry {as the mean brightness difference of the stars we used to construct the PSF, between the 3-px PSF photometry and brightness measured in a 10-pixel aperture}: $0.14$, $0.20$ and $0.21$ in \bb, \vb\ and \ib\ respectively; the \citet{sirianni05} corrections were then applied to correct to infinity. Finally, foreground (galactic) extinction was accounted for using the standard Galactic extinction law \citep[a correction of $A_V\sim0.06$~mag; ][]{schlegel98}. We then restricted the SCC list further by imposing the following criteria: a `sharpness' (a measure of the relative width of a source with respect to that of the PSF) between $-2$ and $2$; a photometric error below 0.3 mag in all bands; and a $\chi$ goodness-of-fit statistic of less than 3. This process follows the first steps of the selection routine tested by \citet{rejkuba05} for stars in resolved populations. 

	From this masterlist, samples of young and old cluster candidates can be drawn according to different luminosity and color cuts. To isolate young clusters one needs to account for the contamination by individual giant stars. This is implemented through a (conservative) cut at an absolute magnitude of \mbox{$M_V<-9$} \citep[standard practice, following the reasoning of][]{whitmore99}, in order to exclude the supergiants that populate the high end of the stellar luminosity function \citep[red supergiants can reach $M_V\simeq-8$~mag;][]{efremov86}. The detection of globular clusters is simplified by their tight distribution in color-space. Therefore, a lower magnitude limit can be enforced while still restricting stellar contamination.

\begin{figure*}
\begin{center}

	\includegraphics[width=\textwidth, angle=0]{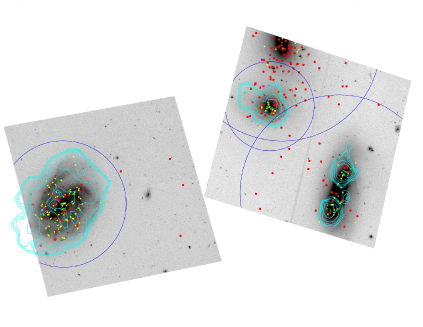}
	
	\caption{Spatial distribution of detected star cluster candidates across the group, color-coded according to age: open green circles and yellow crosses denote nebular and young SCCs respectively, drawn from the final, $M_V<-9$ selected sample. The {red boxes} indicate globular cluster candidates, selected using a fainter brightness limit (see text for selection procedure). The {large blue circles} denote the limits of the galaxy haloes, estimated according to galaxy mass. The proximity of \galb\ and \gald\ and consequent overlap of their haloes makes it virtually impossible to distinguish between their respective GC systems. We have also plotted H\one\ contours in cyan; the levels are 10, 20, 30, 45, 60, 90 and $120\times10^{19}~$cm$^{-2}$. This shows the gas to be contained almost entirely within the individual galaxies (down to the sensitivity limit of the VLA observations), implying a lack of major interactions in recent times. The exception is \gald\ that shows an asymmetric H\one\ distribution. This is potentially the result of stripping from an encounter with \galb\ some time in the past $\sim$Gyr ({revisited in Fig.~\ref{fig:bplusr}}). 
	}\label{fig:regions}
	
\end{center}
\end{figure*}

The application of these stringent criteria  produces a sample of {287} bright star cluster candidates (SCC), of which almost half are situated in and around star-forming \galc. The locations of these sources are shown in Fig.~\ref{fig:regions} and Figs.~\ref{fig:colors}~and~\ref{fig:cmd} show the resulting color-color and color-magnitude diagrams for SCCs. Specifically, the numbers of detected SSCs in galaxies A through D are {50, 28, 133 and 45}, with a further 31 objects in the space outside the galaxies. 

\subsection{Completeness}
In order to test the completeness of our final list of SCCs, we used \texttt{addstar} to add 3000 artificial stars to the image (over the entire field, including superposed on the galaxies) in the magnitude range $24-28$, \ie~$(-9.7, -5.7)$. The average limiting magnitudes for the 50\%\ and 90\%\ recovery rates are (27.5, 26.8), (27.5, 26.3) and (27.1, 26.4) in the \bb, \vb\ and \ib\ bands respectively (after photometric corrections are applied). {Since this calculation represents an average over the entire field, we expect brightness completeness limits for clusters within galaxies. However, this is still far fainter than our $M_V<-9$ cutoff and does not affect the analysis}.

\subsection{Globular cluster candidate selection}\label{sec:gc-selec}
As contamination from luminous
supergiants is not a problem for objects with GC-like colors, we adopt
a fainter magnitude limit to select old GC candidates than we used
for SCCs.  We have chosen a cutoff at $V_{606}=26$, which corresponds 
to $M_V\sim -7.7$, or slightly more luminous than the expected peak in 
the GC luminosity function {(GCLF)} at $M_V\sim -7.4$ \citep[e.g.][]{az98,harris01}. 
At this cutoff, the majority of GC candidates lie above the 90\% 
photometric completeness level in all 3 filters.  Assuming a Gaussian 
GC luminosity function with a peak at $M_V=-7.4\pm 0.2$ and a 
dispersion $\sigma=1.2\pm 0.2$, our cutoff allows us to sample 
$40\pm 8\%$ of the complete GCLF.

We have selected GC candidates (GCCs) according to the color-space
distribution of Milky Way GCs, as featured in \citet{harris96}. More
specifically, we de-reddened the \citeauthor{harris96} catalog
clusters by their listed $E(B-V)$ in the range $[0.00,1.24]$, and then defined a parallelogram
based on the intrinsic $(B-V)$ \textit{vs.} $(V-I)$ color distribution of the MW GCs, 
and then transformed to the ACS filters using the `synthetic'
transformations in \citet{sirianni05}.  All point sources with
1$\sigma$ error bars that overlap the color selection region are
considered GC candidates; they are shown in
Fig.~\ref{fig:regions} and their colors are plotted in 
Fig.~\ref{fig:gcs}.

Much of the contamination in our GC sample will be either from
foreground Milky Way stars or from reddened young clusters.
Predictions from Milky Way star count models \citep[the Besan\c con model of][]{robin03} 
suggest that only $3-4$ foreground stars will appear
in the magnitude/color range for expected GCs in each of the ACS
fields.  Contamination from younger clusters is harder to discern,
particularly in regions of the disks of the spiral galaxies where such
objects should be common (Fig.~\ref{fig:regions}). To quantify, 
the youngest clusters with ages of $\tau \leq 10^7$~yr (the ones most likely 
to suffer from high extinction) would need no more than an $A_V\lesssim1.5$~mag 
to shift to the region occupied by GCs and fall within our criteria. 

Due to the close (projected) proximity of the galaxies in the group
(particularly \gala/B/D), it is likely that the halo GCs in each
system will appear superposed.  In an attempt to quantify the
GCCs in each galaxy, we use the relationship between the galactic mass
and the radial extent of the GC systems in galaxies by \citet{rhode07}.  
To compute the expected size of each halo, we have adopted
the $M/L$ conversions assumed in that work, although we stress the
general conclusions we reach are not dependent on the detailed size of
any given halo.  The expected sizes are 30 kpc for \gala\ and \galb, 20 kpc 
for \galc\ and 14 kpc for \gald; these radii are
shown in Fig.~\ref{fig:regions}.  While we cannot take into account any differences
in the line-of-sight distances to each galaxy, the projected GC
systems of \gala, \galb\ and \gald\ clearly overlap.  We disentangle the
systems in the subsections that follow.  

The resulting GCCs are 21 for \gala, 92 for the overlapping systems of \galb/\gald\ and 29 for \galc.   These will be discussed in more detail in Section~\ref{sec:gcs}.

\section{Additional Observations}\label{sec:moredata}
\subsection{\spit\ IRAC/MIPS imaging in the IR}\label{sec:spit}
The optical imaging was complemented by \spit\ imaging in the {mid-IR} (IRAC, $3.5-8~\mu$m, MIPS $24~\mu$m), presented in J07 and shown here in the bottom panel of Fig.~\ref{fig:finder}. The IRAC band covers stellar photospheric emission and probes the presence of polycyclic aromatic hydrocarbons (PAH) {and hot dust}, while the $24~\mu$m observations trace thermal dust, all sources of emission stimulated by star formation activity. 

The \spit\ images were combined with \textit{JHK$_S$} band observations from \twomass~\citep{2mass} to plot the IR spectral energy distribution (SED) of each galaxy (following J07), presented in Fig.~\ref{fig:seds}. We calculate the slope of the SED within the IRAC bands through a simple power-law fit. This was defined by \citet{gallagher08} as the \airac, and it serves as a measure of star-formation activity. The extent of the flux difference between the PAH features at 8$\mu$m and the stellar-blackbody emission at 3.5$\mu$m leads to a positive gradient {(blue color)} in quiescent environments, while star formation registers as a negative slope {(red color)}. The steepness of the slope scales with the intensity of the star formation.

\subsection{VLA 21-cm observations: derivation of gas mass and column density}\label{sec:vla}
We also make use of radio wavelength observations, in order to investigate and characterize the state of neutral hydrogen gas in the system. While H\one\ observatinos of HCG~7 have been presented in the past \citep{huchtmeier97,verdes01,borthakur09}, these {single-dish} data covered the system as a whole and not the member galaxies on an individual basis. We used archival VLA H\one\ data, obtained in the C and D array configurations, to map the {column} density of H\one\ across the system and to derive the masses of all member galaxies. 

The reduction and analysis of these data will be presented in full in Heiderman~\etal~(in preparation; hereafter H10); we will refer to this work in the following sections. In brief, we used standard techniques and procedures in the Astronomical Image Processing System (AIPS; see Rupen et al. 1999) to reduce the data.  The data contained major interference issues, which were carefully corrected before calibration and bandpass-correction.  The data were continuum subtracted by making a linear fit to the visibilities, and then combined in the UV plane.  The C+D combined data have a spatial resolution and channel separation of $19.4^{\prime\prime} \times 15.2^{\prime\prime}$ and $21.3$~\kms. The on-source integration time totaled $\sim$~7 hours and the number of combined spectral channels were 54.  This corresponds to a $1-\sigma$ rms noise of 0.3 mJy beam$^{-1}$ and a
$\sigma(N_{\rm H{\scriptsize \sc I}})$ 
column density sensitivity of 2.4~$\times 10^{19}$ cm$^{-2}$ per channel over the beam size.

We used a moment zero map to derive a mapping of column densities and an estimate of the H\one\ mass contained in each galaxy. 
We follow a simple transformation of radio intensity ($S_\nu$) to brightness temperature ($T_b$) and then to neutral hydrogen column density (\nhi) and mass (\mhi):

$$	T_b = S_\nu \times c^2 / (2\,\nu^2\,k)~~\textrm{Kelvin}	$$
$$	N_\textup{\scriptsize H\one} = 1.8224\times10^{18} \times \int T_b\,dv~~\textrm{cm}^{-2}	$$
$$	M_\textup{\scriptsize H\one} = 2.36\times10^5 \times d^2 \int S_\nu\,d\nu~~\Msun	$$
where $c$ is the speed of light in vacuum and $k$ is the Boltzmann constant. 

The derived column densities are shown as green contours in Fig.~\ref{fig:regions} and masses of galaxies \gala, \galc\ and \gald\ were calculated as \mbox{$\log$~\mhi~$ = 9.0,~9.5,~8.9$ \Msun}; no H\one\ gas was detected in the vicinity of \galb.

\subsection{Las Campanas wide-field imaging}\label{sec:campanas}
The imaging coverage is extended through our wide-field mosaic imaging of the group and its surroundings in the $B$ and $R$ bands. These images were obtained with the Direct CCD Imager on the Du Pont 2.5-meter telescope at Las Campanas, Chile on the nights of 2007 Oct 3/5 under good seeing conditions (approximately 0\farcs75), as part of a project covering all 12 HCGs in our sample. The exposure times were 180 and 120 seconds with the $B$ (JB 3008) and $R$ (KC-R 3010) filters respectively, allowing for {point sources} to be detected down to roughly 24 mag (in either filter) at \mbox{S/N~$=5$}. 

These images are used to look for signatures of ancient interactions, a subject we discuss in full in Section~\ref{sec:bplusr} (and Fig.~\ref{fig:bplusr}). Their primary purpose was, however, to provide a list of candidate dwarf galaxies in HCG~7 for spectroscopic follow-up. This program is ongoing with the Hydra spectrograph on the CTIO 4-meter telescope and will be presented in full once concluded. A brief account of the data acquisition and reduction is offered in the following section. 

\subsection{CTIO Hydra multi-object optical spectroscopy}\label{sec:hydra}
The dataset is completed by optical spectroscopy of candidate dwarf galaxies selected through the Las Campanas imaging campaign presented in Section~\ref{sec:campanas}. This ongoing spectroscopy project will eventually cover all compact groups in our multi-wavelength dataset and will be presented in full in a later work. 
{The HCG~7 data-set} was obtained with the Hydra multi-fiber spectrograph on the Cerro Tololo Interamerican Observatory (CTIO) 4-m telescope. Observations were carried out between 3-10 August 2008 with a total on-target integration time of
$\sim 13.5$ hours, {with various fiber configurations, covering a different set of targets in each exposure}. We used grating KPGL2 with the GG 385 blocking filter (position 2)
to avoid order contamination. The SITe CCD was binned by 
a factor of 2 in the dispersion direction to reduce readout noise.
The resulting spectra cover the range $\sim 3000 - 7800$\AA\ at a dispersion of
2.3~\AA\,px$^{-1}$. {The resolution, measured as the full-width at half-maximum of a sky-line at $\sim5000$~\AA, is 6.1~\AA}. Wavelength calibration was facilitated through exposures with the
He-Ne-Ar penray lamps. 
We took twilight flats to compensate for fiber-to-fiber
throughput variations and `milky flats' to remove the 
two-dimensional instrumental spectral shape\footnote{{This is 
a daylight flat that is taken in order to illuminate wavelengths that register very faintly under the artificial light of calibration lamps. As the daytime sky is ÔwarmerÕ than a lamp, we use that to obtain a smooth flat-field division across the broad wavelength range covered.}}.
We used the Hydra-specific software {\tt hydraassign} to
construct fiber configurations for targets, guide stars and 
blank sky observations. Guide stars were selected
from the Sloan Digital Sky Survey \citep[SDSS,][]{sdss} that fully covers the HCG 7 field.
Each configuration
included at least 12 guide stars and 30 blank sky observations 
and a list of targets tiered according to magnitude, in order to 
optimize exposure times. 
We thus produced a set of five configurations that included progressively
fainter targets, with the aim of obtaining a complete, flux-limited set of 
observations for our targets. We estimate that our observations for
HCG 7 targets are complete down to $R\sim 20.5$. 

Data reduction was carried out using the standard 
IRAF {\tt ccdred} package as well as the
the dedicated IRAF package {\tt hydra}. We used 
LACOSMIC \citep{lacosmic} to eliminate
cosmic rays in individual exposures before spectral extraction.
After spectral extraction, we used custom-written IRAF scripts
to combine all data and error spectra at the same configuration, 
weighting them by their variance.

In HCG~7 we identify a single additional member with high-confidence, a compact galaxy to the northwest of \galb, at a projected distance of $\sim30$~kpc. This one source results from a list of hundreds of candidates down to $R\sim20.5$. Its spectrum is shown in Fig.~\ref{fig:spectra}. 

We also used this program to place fibers on the four main galaxies, also shown in Fig.~\ref{fig:spectra}. {The 2$^{\prime\prime}$ width of the Hydra fiber translates to $\sim0.8$~kpc at the distance of HCG~7, 
therefore the spectra cover the central regions of the galaxies}. A quick inspection {shows features inconsistent with 
their previous designation} of morphological types: \gala\ shows very bright emission in \ha\ and some in \hb, features not typical of an Sa; the spectrum of \galc\ is not {dominated by the blue continuum}, as one would expect from an Sc; while \gald\ features \ha\ emission, {it is embedded in a deep stellar absorption line, indicative of a post-burst spectrum}. The spectroscopic classification of HCG~7 galaxies will be revisited in Section~\ref{sec:seds}, where we present a combined analysis of their spectra and optical-to-mid-infrared spectral energy distributions.

\section{The present state of HCG~7: gas content and star formation activity}\label{sec:s-pops}
The current consensus in star formation research suggests that the vast majority of star formation occurs in a clustered fashion. Owing to this fact, the processes of star and cluster formation are inextricably linked.  The cluster population of a galaxy therefore provides information on its current state and recent star formation history.   Here we will provide an interpretation of the photometric properties of the detected star cluster candidates in HCG~7. We combine this information with overall galaxy properties to draw implications on the evolution of this compact group.

\subsection{Star clusters}
We compared the colors of all SCCs to models of simple stellar populations (SSP) by \citet{bc03}. These models only predict properties of the stars, and do not incorporate either continuum or line emission from the gas that surrounds very young clusters. We assume a metallicity of $\frac{1}{5}$~\Zsun\ for all four galaxies, following the mass-metallicity relation observed in star-forming galaxies \citep{ellison09}. Unfortunately, as our analysis lacks $U$ band imaging, we are unable to break the age-extinction degeneracy characteristic for SSPs with ages $\tau\lesssim1$~Gyr. The model track runs virtually parallel to the extinction vector, hindering the distinction of old clusters from reddened younger clusters based on colors alone. However, the goals of this study can be met through a qualitative age-dating process: we are investigating distinct epochs of star formation and those can still be identified. 

\begin{figure*}
\begin{center}

	\includegraphics[width=0.7\textwidth]{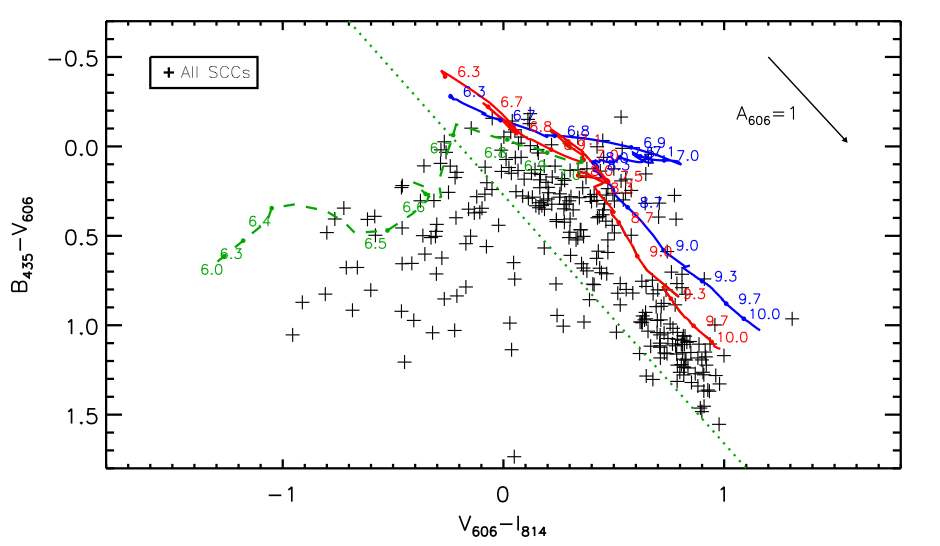}\\
	\includegraphics[width=0.7\textwidth]{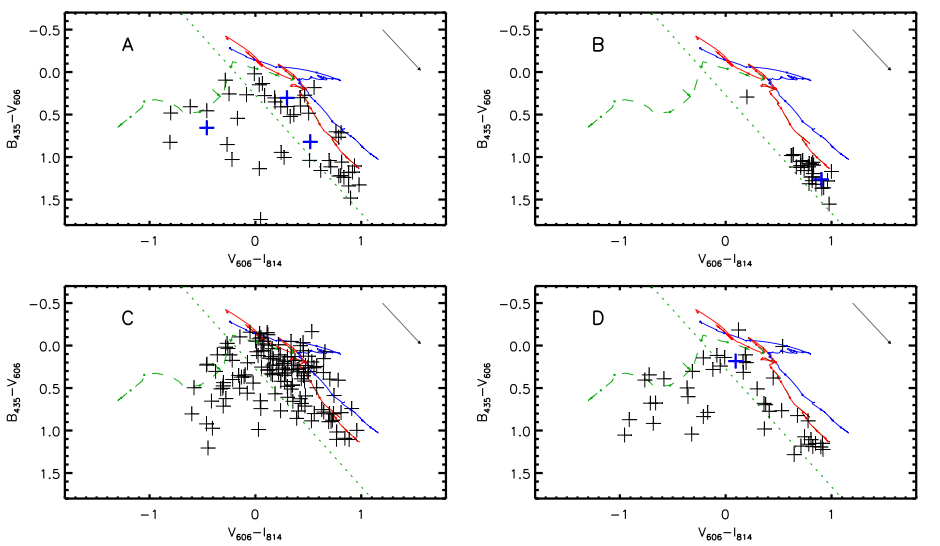}\\
	
	\caption{\small \bb$-$\vb\ \textit{vs.} \vb$-$\ib\ colors of the entire sample of SCCs ({\bf top}) and {divided by regions} ({\bf bottom}), plotted on top of $\frac{1}{5}$~\Zsun\ (solid red line, bottom) and \Zsun\ (solid blue) BC03 model tracks. The dashed green line shows an evolutionary track that accounts for nebular emission (SB99), as suited to the very youngest clusters that have not yet expelled their natal gas. All data-points that lie to the left of the dotted green line are consistent with the nebular tracks, given an amount of extinction of no more than $\sim1$~mag. SCCs (\ie\ those with $M_V<-9$) are denoted by dark crosses; we indicate an extinction vector of length 1~mag in $V_{606}$. In the plots of individual galaxies, blue (bold) crosses mark the very brightest sources with $M_V<-11$, while we denote the entire sample as dots for comparison. A typical error bar is $\lesssim0.1$~mag in either direction. All three spiral galaxies show signs of star formation over a Hubble time, as indicated by the spread of cluster colors about the model tracks. The lenticular \galb\ only shows signs of cluster formation at ages greater than a few Gyr. \normalsize
	}\label{fig:colors}
	
\end{center}
\end{figure*}

\begin{figure*}
\begin{center}

	\includegraphics[width=0.7\textwidth]{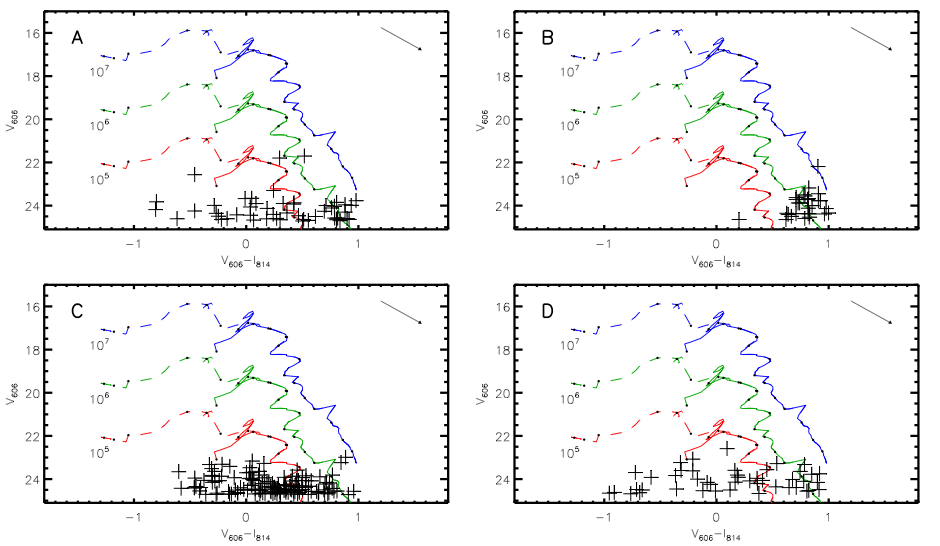}\\
	\includegraphics[width=0.7\textwidth]{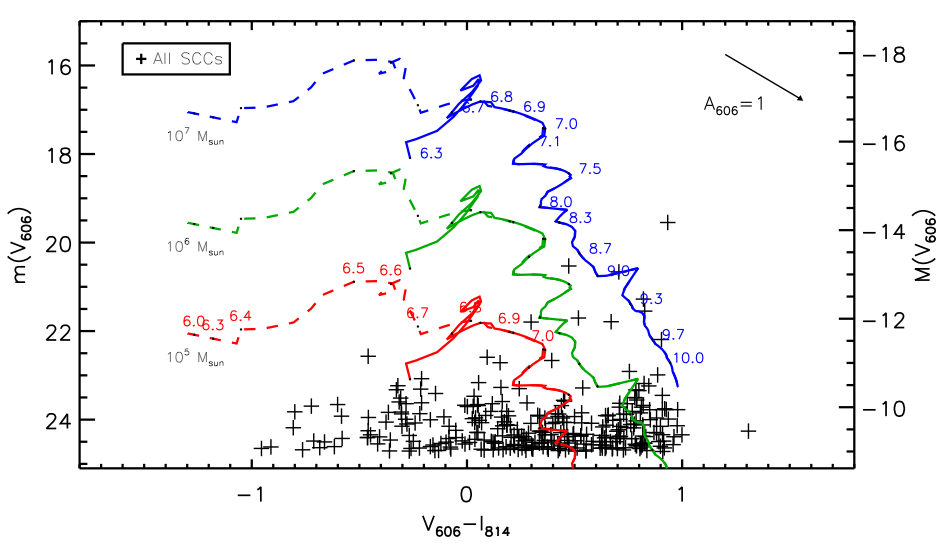}\\
	
	\caption{\vb\ \textit{vs.} \vb$-$\ib\ color-magnitude diagrams for the full sample ({\bf top}) and {divided by regions} ({\bf bottom}). We have provided the same model tracks as in Fig.~\ref{fig:colors} for three different cluster masses (solid lines), with the `nebular' tracks plotted as dashed lines. From this it follows that all cluster candidates have high masses. {This is expected, as the large distance imposes a very faint detection limit for low-mass clusters}. The SCCs in the three spiral type galaxies display similar distributions in these plots, while in \galb\ all SCCs are tightly concentrated. This is related to the narrow distribution of globular cluster colors. 
	}\label{fig:cmd}
	
\end{center}
\end{figure*}

	This analysis is presented in Figs.~\ref{fig:colors}, \ref{fig:cmd}, where we show the $B-V$ vs $V-I$ color plots and corresponding color-magnitude diagrams. The large plots show the collective properties of the HCG~7 cluster population, while the smaller panels pertain to individual galaxies. Each of these panels shows the `footprints' of all SCCs in HCG~7 as dots, with individual galaxy members denoted by crosses. Blue (thick) crosses indicate the very brightest sources, those with $M_V<-11$. 
	
	Crucially, the very youngest clusters can be distinguished by their strong nebular emission, caused by the ionization of their surrounding gas. This is either the residual gas of star formation that is yet to be expelled, or gas in the general vicinity of a cluster-forming region, both indicative of a young age (although line of sight effects are possible in the latter case). The result of this is emission in \ha\ and [O\three], both covered in the \textit{F606W} filter ($z_\textup{\scriptsize HCG~7}\simeq0.014$, \ie\ a wavelength-space shift of $\Delta\lambda\simeq90$~\AA). 
	
	In order to incorporate that information, we obtain the strength of the \ha\ and H$\beta$ emission lines from Starburst99 models \citep[][SB99]{sb99} with the same metallicity and IMF. As SB99 does not include the [O\three] line, we defined it through the ratio of log([O\three]/H$\beta$). We used the highest value for this diagnostic of 0.7 \citep[as found in the KISS sample of low-mass star-forming galaxies; ][]{salzer05} to estimate the temporal evolution of the [O\three] emission line strength. {This is, of course, a qualitative treatment of the problem at hand: the relative strength of [O\three]/H$\beta$ is environment- and metallicity-dependent and mildly varies from one cluster to the next, according to the properties of the gas \citep{oestlin03}. There are more lines that should be considered for a robust treatment of this issue, such as [OII]. 
	
	In effect, our treatment is sufficient to separate `nebular' sources:} young clusters with nebular emission can be distinguished from older clusters by roughly dividing color-space into regions. First, we draw a line parallel to the extinction vector, crossing the nebular model track at $\log\tau=6.7$ (the dotted green line in Fig.~\ref{fig:colors}). We consider any source to the left of this line to have nebular emission, \ie\ to belong to a cluster-forming region, and therefore assign an age of $\lesssim 10$~Myr. {We must caution at this point that \ha\ emission does not prove that a source is younger than 10~Myr, but solely that it is surrounded by ionized gas. Such chance alignments have been found to be a common occurrence in nearby environments by \citet{gelys07a} and \citet{isk09a}, as deduced through the spectroscopic examination of the absorption (stellar) and emission (nebular) line signatures. This might arise from the limited spatial resolution of the ground-based observations of the aforementioned works. The issue is treated in \citet{whitmore10}, where it is found to have a limited impact in high-resolution, \hst-based studies. In addition, the absence of high -resolution $U$-band imaging disqualifies precise cluster age-dating on an individual basis. A cluster that is found in an \ha\ bright region is most likely to be no more than a few tens of Myr old, an age space that cannot be resolved by our diagnostics.}

	Another region is drawn to distinguish GC candidates (GCC). This rectangle is based on the observed colors of Milky Way GCs \citep{harris96} and is indicated on Fig.~\ref{fig:gcs}. Any source not belonging to one of these regions is considered a candidate young massive cluster (YMC). Note that we also apply a color-cut at \bb$-$\vb$<1.5$~mag or \vb$-$\ib$<1$~mag to exclude foreground stars. {This process was outlined in \S~\ref {sec:gc-selec}}.
	
The vast majority of young clusters (nebular included) are located in or around the actively star-forming spirals. In Fig.~\ref{fig:regions} we show the spatial distribution of all cluster candidates, coded in green, blue and red for nebular, young and old respectively.
	As expected, the lenticular \galb\ is found to host the largest population of old clusters, 
	In fact, \galb\ hosts no detectable candidates aged less than a few Gyr, implying that star formation slowed down significantly long ago. 

{The color-color plots of the spiral galaxies reveal an apparent lack of clusters at $\tau\sim 1$~Gyr}. This is a common feature in populations of star clusters and arises primarily because of the evolutionary fading of stellar aggregates: with stellar evolution gradually extinguishing the most luminous stars, the mass-to-light ratio of SSPs increases steadily with time. If $M/L\propto\tau^\alpha$, then $L\propto M/\tau^\alpha$, meaning that a constant magnitude cutoff will restrict our sample to higher mass with increasing cluster age. This results in fewer clusters being selected once we cross the Gyr mark. In addition, the lack of $U$-band observations inhibits the distinction between an old cluster and a reddened younger cluster. It takes less than two magnitudes of extinction to displace a cluster from $\log\tau\sim7$ to $\log\tau\sim10$. In that way, a young cluster in a dusty star forming region may mimic a globular.

\subsection{Cluster complexes}
Star clusters generally do not form in isolation. Like the stars they contain, clusters represent a level of hierarchical structure formation within galaxies. It is believed that the majority of clusters form in complexes that cover a large range of masses and sizes. This connection was explored by \citet{elmegreen83}, who found that spiral arms provide the appropriate conditions for the largest complexes to form. As these gas agglomerates can reach masses of $\sim10^7$~\Msun, the high end of their mass distribution can be observed to great distances. In nearby environments the size and mass distributions are better sampled, and relations between their physical properties can be pursued to greater detail. \citet{bastian05b} found them to obey a mass-radius relation of the form $M\propto R^{2.33\pm0.19}$, a link that is communicated between complexes and GMCs \citep[$M_\textup{\scriptsize GMC}\propto R_\textup{\scriptsize GMC}^2$, ][]{solomon87}. 
	
	As complexes represent such a fundamental part of the star formation process, they merit an in depth study. At large distances, however, their detection is non-trivial: the spatial resolution limit of even \hst\ gives rise to sample contamination, especially in inclined spirals, where chance alignments are common. 
	
\begin{figure*}
\begin{center}

	\includegraphics[width=0.8\textwidth, angle=0]{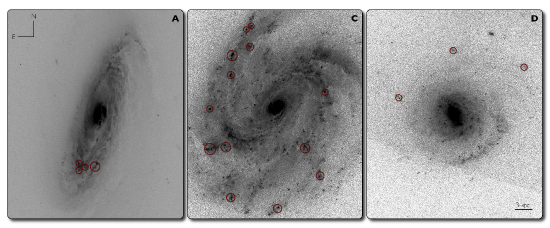}
	\includegraphics[width=0.8\textwidth, angle=0]{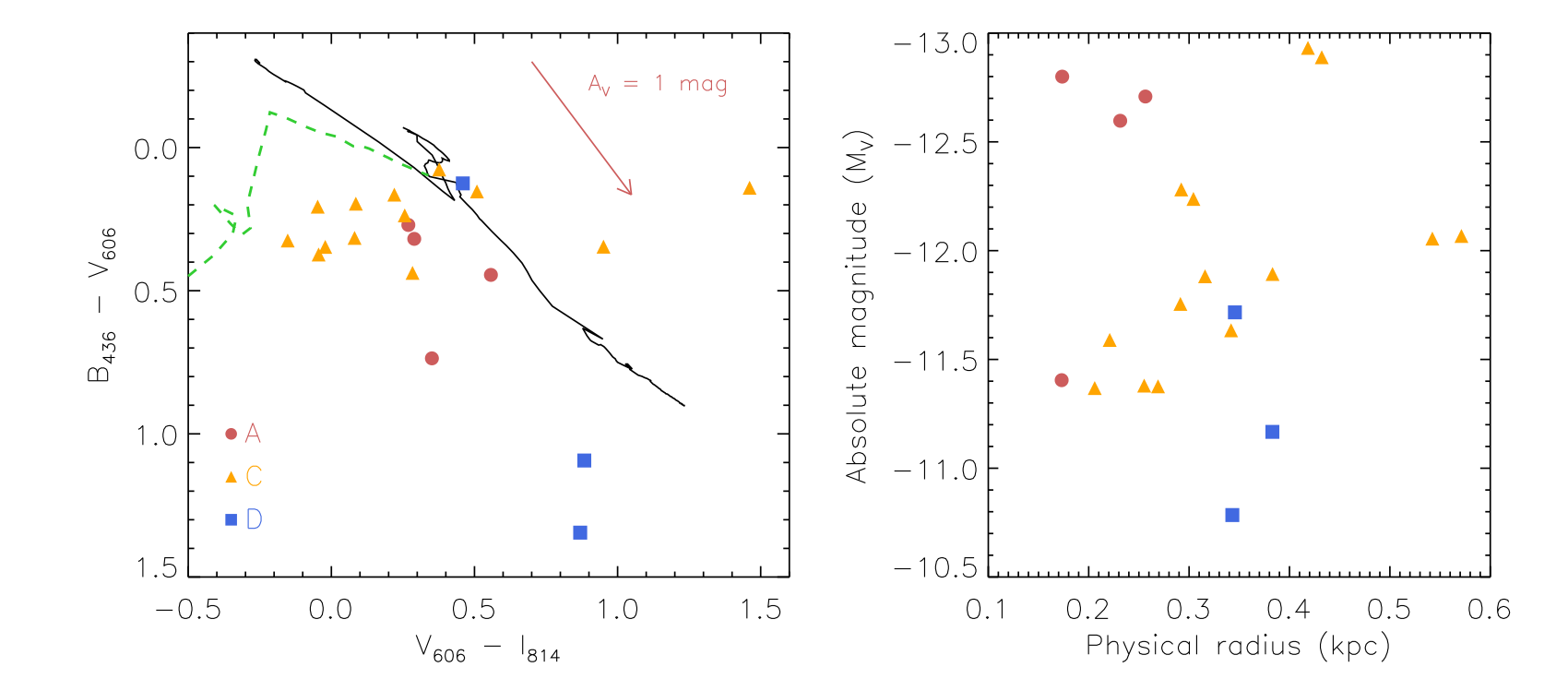}
	
	\caption{Star cluster complexes in HCG~7 (\hst\ \vb\ images). In the {\bf top} panel we have indicated their locations in each of the three star forming galaxies (presented to scale). These structures represent the highest level of the star formation hierarchy within galaxies. They contain a number of star clusters and represent more than a single epoch of cluster formation. In accord with this view, we find the complexes to have colors, and therefore ages, consistent with young stellar populations -- as presented in the {\bf bottom left} panel. 
	Their absolute magnitudes are plotted against their sizes in the {\bf bottom right} panel; the magnitudes scale approximately with galaxy {brightness, \vb$\,=\,$}13.4 (\gala), 13.8 (\galc) and 15.4 (\gald). Complexes in \gala\ are difficult to measure because of inclination effects. 
	}\label{fig:complexes}
	
\end{center}
\end{figure*}

	In this work we studied of the properties of 21 complexes -- 4, 14 and 3 in \gala, \galc\ and \gald\ respectively, shown in Fig.~\ref{fig:complexes}. We selected them by eye (the large pixel scale at this distance impedes automatic detection), roughly measured their sizes as the extent that lies about 10 sigma above the background, 
and compared their colors (after background subtraction) to SSP models in order toto obtain age estimates. While the possibility of contamination is quite high in the inclined \gala, the sizes and colors of the detected groupings are consistent with the distributions established in the local universe \citep{elmegreen99}.  More specifically, most complexes have colors indicating nebular emission features, as expected by their high gas content and young ages; that is to say, they stray to the left of the model track, as do gas-enshrouded clusters (Fig.~\ref{fig:complexes}). Two of the three complexes in the inclined spiral \gala\ appear to be heavily reddened ($A_V\sim2.0$~mag), in tune with their locations {in the galaxy}. 
	
	In all, the {sizes and colors of} complexes in the three galaxies are consistent with quiescent systems found locally (see above references). This is to be expected, as the size of complexes is only dependent on that of the parent galaxy. The fact that we detect young star clusters of various ages associated with these complexes supports their role as the primary sites of star cluster formation (ongoing until depletion) and therefore the hierarchical nature of star formation in galaxies.

	\subsection{Star formation rates and stellar masses}\label{sec:sfr}
	Key to deciphering the present state of the system are the star formation rates of the individual galaxies. {For this}, we adopt the measurements of \citet[][hereafter T10]{tzanavaris10}. That work combines the \spit\ $24~\mu$m described in Section~\ref{sec:spit} with $2000~$\AA\ photometry from the \textit{Swift} ultraviolet and optical telescope \citep[UVOT;][]{roming05} to derive the `true' star formation rate. {The two wavebands capture} different stages of star formation: the IR probes star-forming clouds, where protoclusters can be found deeply embedded in their natal clouds \citep{ladalada03}. The high column densities of dust in these clouds absorb the UV photons emanating from protostars, become thermally excited and re-emit the radiation at longer wavelengths. The timescale for this phase (the age of a protocluster) is $\tau_\textup{\scriptsize IR}\lesssim1$~Myr, thus the mid-IR SFR measures the current SF activity. At a timescale of $\sim100$~Myr, the clouds have been expelled by the mechanical feedback of stellar winds and supernov\ae\ and the UV photons, so richly emitted from the photospheres of young stars, are free to escape along many lines of sight. The SFR derived from the UV is therefore telling of a more prolonged era of star formation and is appropriate for studying quiescent environments. It does, however, miss information of current activity ($\tau\lesssim10$~Myr) that is dust-enshrouded and therefore emerges at longer wavelengths. 

	T10 calculate the rates of \gala/\galb/\galc/\gald\ as \mbox{$3.88\pm0.47$}, \mbox{$0.23\pm0.02$}, \mbox{$2.06\pm0.17$} and \mbox{$0.43\pm0.04$}~\Msun\,yr$^{-1}$. Given the variation in galaxy sizes, we can relate the star formation rates to the stellar mass to get a better handle on the current star formation activity in the HCG~7 galaxies. T10 divide by the stellar mass of each galaxy, as derived from 2MASS $K_S$-band photometry (see J07, Table~\ref{tab2}), to obtain specific star formation rates (sSFR) of $23.46\pm4.25$, $2.56\pm0.42$, $27.33\pm4.34$ and $11.89\pm1.90\times10^{-11}$~yr$^{-1}$. The relation between disk-averaged star formation rates and \mhi\ is in all cases consistent {with the correlation found in} \citet{kennicutt98}. 
	
	Comparing to the SINGS dataset \citep{sings}, these rates suggest the three disk galaxies are moderately star forming, with star formation rates at the high end of the quiescent  regime, given their morphologies, sizes and stellar content. The lenticular \galb\ is also in tune with its morphological counterparts, again, at the high end of the SFR distribution. Comparing to the sample of nearby galaxies treated by \citet{leroy08}, the sSFRs are comparable to those of {star-forming} galaxies in that sample -- although because of the low numbers, this is not a statistically significant comparison. This indicates relatively high star formation activity throughout the group. Finally, we compare to the AMIGA sample of isolated galaxies \citep{amiga1,amiga2}, which, unfortunately, does not have associated stellar mass information. A comparison based on morphological type alone places the SFRs one sigma above the median {values of 0.72 and 1.09 \Msun\,yr$^{-1}$ for Sa and Sc types respectively, where the corresponding standard deviations are 1.22 and 2.81 \Msun\,yr$^{-1}$}.

All the values quoted in this section are listed in Table~\ref{tab2}.

\subsection{Detection of compact galaxies}\label{sec:dwarf}
Our spectroscopic campaign to identify group members further from the four main galaxies (Section~\ref{sec:hydra}) yielded a single high-confidence detection. This is a compact galaxy to the northwest of \galb, apparently associated with a stellar streamer (Section~\ref{sec:bplusr}) to which we refer as \galx. It was detected in the SDSS, where it cataloged as as J003915.46+005633.2. Its spectrum is shown in Fig.~\ref{fig:spectra} and its radial velocity is measured as 4118~\kms\ (determined through the C\,\two~H \&\ K lines). The spectrum is characteristic of an old elliptical galaxy, with no discernible emission features, as also confirmed by the SDSS spectrum. Its redshift gives rise to a luminosity distance of $D_L=57.8$~Mpc and therefore an absolute blue magnitude of $M_B=-15.2$~mag. 

Its brightness and spectrum are consistent with a dwarf elliptical galaxy. Its spectrum will be revisited in a later work, where we will be examining the dwarf galaxy populations of all HCGs in our multi-wavelength sample.

\subsection{Searching for optical signatures of ancient interactions}\label{sec:bplusr}
\begin{figure*}
\begin{center}

	\includegraphics[width=0.4\textwidth, angle=0]{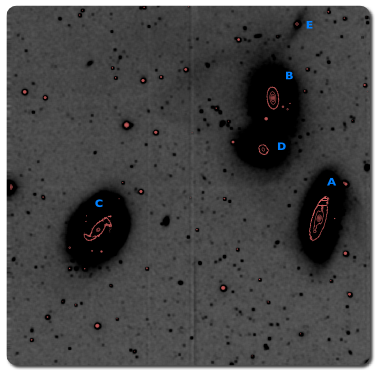}~\includegraphics[width=0.4\textwidth, angle=0]{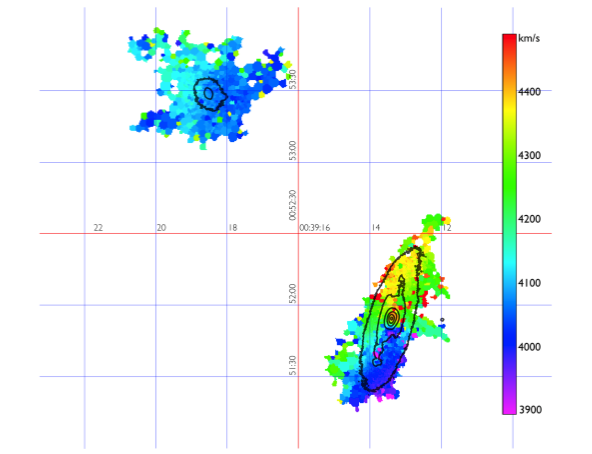}
	
	\caption{
	{\bf Left: }Smoothed, co-added $B+R$ image of HCG~7 (Las Campanas 2.5m; Section~\ref{sec:campanas}), presented at high contrast to enhance faint signatures of diffuse IGM components. We have overplotted $R$-band contours in a logarithmic scale, to demonstrate the main structure of each galaxy. We use this image to search for evidence of past interactions. While we find no compelling evidence of ancient encounters between the four main galaxies, we detect a streamer connecting \galb\ to the compact galaxy \galx\ (Section~\ref{sec:dwarf}), {as well as} a very faint structure to the east and north of \galb. More details are presented in Section~\ref{sec:bplusr}. {\bf Right: } Fabry-P\'erot velocity map, adapted from \citet{torres09}, covering galaxies \gala\ and \gald. {The contours are the same as on the left panel}. There is evidence for a streamer to the north of \gald, \ie\ the direction of its close neighbour, \galb. There is also apparently a velocity-consistent structure to one side of \gala, which also registers in H\one\ (Fig.\ref{fig:regions}). We interpret this collimated stream as a galactic wind emanating from the nucleus of \gala~(Section~\ref{sec:seds}). 
	}
	\label{fig:bplusr}
	
\end{center}
\end{figure*}

Interactions and mergers lead to the redistribution of material about a system. In search of traces left behind from past interactions, we co-added the $B$ and $R$ band wide-field images of HCG~7 (presented in Section~\ref{sec:campanas}) and smoothed the resulting image with a $15\times15$ pixel boxcar. These two bands provide a good tracer of star formation, as they cover young stars and \ha\ emission respectively; $R$ band also covers light from old stars. We present the smoothed, co-added image at high contrast in Fig.~\ref{fig:bplusr}. 

At first glance, there is no compelling evidence of major interactions among the four member galaxies; \galc\ shows no signs of disturbance. However, we do find faint stellar features in the system, all related to \galb: a streamer to its north, seemingly connecting it to the compact galaxy \galx, some $30$~kpc in projection. Based on its {marginal} detection in $B$-band alone, we propose that star formation along this stream has long been extinguished. We find the color of this feature to be approximately $(B-R)=1.1$, indicative of an old age. It compares to a BC03 SSP of age $\sim2$~Gyr. This is an uncertain measurement, due to the {marginal} detection in the $B$-band, thus we consider this age a lower limit: if the stars formed in a tidal feature, then the passing of \galx\ about \galb\ occurred at that time. Otherwise, if the streamer consists of pre-existing stellar material, the youngest component of the mixed-age population has that age. In the following section we consider the distribution if H\one\ gas about HCG~7, in search of gaseous counterparts of the discovered streamers. 

We also see the faint signature of an arc about \galb, extending to the east of the galaxy and possibly delineating the orbit of \galx\ or \gald.

In turn, the central regions of \gald\  present a symmetric light profile. Its spiral structure, however, is asymmetric, with the arm facing the side of \galb\ appearing very diffuse. This may provide evidence of a past interaction between the two systems. Unfortunately, the stellar haloes of the two galaxies are largely overlapping (Fig.~\ref{fig:bplusr}), therefore we cannot look for tidal features in the space between them. 

Finally, \gala: while its isophotes \gala\ appear slightly irregular, this could be due to the imprint of its spiral structure at high inclination (this issue is revisited in Section~\ref{sec:seds}). The space between \gala\ and the \galb/D complex may possibly feature another faint signature, although, again, this is enhanced by -- if not entirely due to -- the superposition of the two haloes. 

All of this information is telling of interactions. Given that collimated structures are short lived, especially in a complex tidal field, these interactions most likely happened in the past $\sim$Gyr.

\subsection{The gaseous component of HCG~7}\label{sec:gasmass}
We now turn our attention to the {cold} gas content of this galaxy group and analyze the available VLA 21-cm H\one\ observations. We find the gas to be contained entirely in the individual galaxies down to the $2.4 \times 10^{19}$ atom\,cm$^{-2}$ sensitivity limit of our dataset (derived in H10). This limit translates to a gas {surface} density of \mbox{roughly 0.19~\Msun\,pc$^{-2}$}. This apparent absence of H\one\ in the intra-group medium implies a lack of major interactions between the four galaxies for at least one crossing time: $\tau_\textup{\scriptsize cross} = R/\sigma \sim 60~\textup{kpc}/100~\textup{\kms}\sim0.6$~Gyr. Note that the lack of three-dimensional velocity information and the small group membership make both parts of this fraction uncertain to a factor of two, so a $\tau_\textup{\scriptsize cross} \sim 2$~Gyr is plausible. 

This timeframe is in accord with our understanding of dispersion timescales for tidal features: a suitable empirical upper limit of the ages of star clusters in such environments is $\sim500$~Myr. {This value reflects the highest ages found for tidal clusters in previous studies \citep[\eg][]{schweizer98,gelys07a}. This empirical upper limit might be due to the fading of clusters with age, rather than their dispersal or destruction. Even in this case, our reasoning would still hold: HCG~7 clusters are subject to the same observational limits, therefore we can assume that 500 Myr is the age of the oldest cluster we expect to be able to detect -- be it due to fading or absence}. Furthermore, \citet{hibbard99} detected an H\one\ tidal structure in Arp~299 with an age of $\sim750$~Myr with no optical counterpart. This implies that the gaseous component of tidal tails is more prevalent than the stars -- as displayed, \eg, in a plethora of multiple systems in the H\one\ \textit{Rogues Gallery} \citep{rogues}. 

Additionally, the H\one\ disks do not extend beyond their optical counterparts, an effect quite unusual outside the cores of galaxy clusters. This attribute may indicate that H\one\ preprocessing has occurred within the galaxies, seemingly in the absence of major external influence. The exception here is \gald, which presents a slightly asymmetric H\one\ distribution, reflected in its spiral structure. It is worth noting that the asymmetry is pointing towards \galb, however there are none of the tell-tale signs of a close encounter, such as a tidal bridge or young star clusters in the region. It is interesting to compare to the results of the Fabry-P\'erot interferometry presented by \citet{torres09}, who find evidence of a gaseous streamer directed at \galb. In addition, there is a clump of H\one\ below the nucleus of \gala, co-spatial with {an H$\alpha$ streamer} detected in the Fabry-P\'erot map ({adapted in Fig.~\ref{fig:bplusr}}). These two pieces of evidence strongly suggest an outflow from the nucleus of \gala, a very likely result of a starburst; {we confirm this interpretation with our spectra in Section~\ref{sec:seds:a}}. 

	In \ref{sec:vla} we derived the H\one\ masses for \galacd\ as \mbox{$\log$~\mhi~$ = 9.0,~9.5,~8.9$ \Msun}, with a null detection in the lenticular. H10 estimate an upper limit by properly accounting for the velocity width of an S0 galaxy and the RMS error of the observations of $\log M_\textup{\scriptsize H\one}^{\scriptsize\galb}<8.06$~\Msun, a reasonable value for an evolved galaxy. These values are collected and contrasted with stellar masses and star formation characteristics in Table~\ref{tab2}. 
	
	The resulting total H\one\ mass of {$M_\textup{\scriptsize H\one}^{tot}=9.72$~\Msun} is consistent with the single-dish measurement by \citet{verdes01}\footnote{\citet{verdes01} calculate $M_\textup{\scriptsize H\one}^{tot}=9.68$~\Msun\ using a single dish measurement of the entire group; this small discrepancy is reasonable, considering that the interferometer is insensitive to large scale H\one\ emission.} and is approximately one third of the average found in galaxy groups \citep{haynes84}. 
		
	Naturally, hydrogen in galaxies is not only found in its neutral form, which is most likely inert in the star formation process \citep{kennicutt07, bigiel08, blanc09, krumholz09}.	As stars form out of cool molecular gas, it is interesting to know the amount of H$_2$ in each of the star-forming spirals. We draw these values from the study by \citet{vm98}, where \mhtwo\ is derived through the amount of molecular CO (through observations of the \mbox{$J=1\rightarrow0$} and \mbox{$J=2 \rightarrow 1$} transition lines). The values for \galacd\ are $\log\,$\mhtwo$ = 9.7, <9.2, 8.4$, where the value for \galc\ is an upper limit. We note that the CO-to-H$_2$ conversion factor (\xco) is far from certain -- derived values range between $\sim2-3\times 10^{20}$ with associated uncertainties of  $\sim0.7-1.5 \times 10^{20}$ -- therefore we treat these values with caution. 
	
	The above gas masses can now be used as a basis for an estimate of the timescale for depletion of the gas reservoir -- both H\one\ and H$_2$. Following a rough assumption that i) the current SFRs will be sustained until the depletion of all gas and ii) the H\one\ reservoir is gradually being converted into H$_2$ and thereafter into stars ({with an eventual efficiency of 100\%}), the timescale for the depletion of H\one\ gas in HCG~7 is of the order of 1~Gyr: \mbox{$\log\,$\tquench\,$=8.7,~9.2,~9.3$~yr} for \galacd. This provides another interesting comparison to the \citet{leroy08} sample, where the median value is \mbox{$\log\,$\tquench$\,= 9.7$~yr}, \ie\ a few Gyrs longer. When considering the H$_2$ values inferred by \citeauthor{vm98}, the timescale for the H\one\,+\,H$_2$ reservoir becomes an order of magnitude longer in the case of \gala: \mbox{$\log\,$\tquench$\,=9.2, 9.4, 9.4$~yr} for \galacd~(an upper limit in the case of \galc). This is due to the large H$_2$-to-H\one\ mass ratio of \gala\ ($\sim5$) that signifies that most of the gas is currently fueling star formation. We will revisit this intriguing find in Section~\ref{sec:seds}. Unfortunately, given the uncertainty in the \xco\ conversion factor, these values have a potentially large systematic uncertainty. We have omitted the contribution by He gas in the above calculation, as it is outweighed by the uncertainty in \xco.

\begin{table*}
\begin{center}
\caption{Derived photometric and other information.}\label{tab2}
\begin{tabular}{cr@{~$\times$~}lccr@{~$\pm$~}lr@{~$\pm$~}lccrrr} 
\tableline
\tableline
ID &			
\tmult{Dimensions}&	
$\log M_*$ &		
$\log M_{\textup{\scriptsize H\one}}$ &	
\tmult{SFR} &		
\tmult{sSFR} &		
$\log\Sigma_\textup{\scriptsize SFR}$\tablenotemark{a} & 
$\log\Sigma_\textup{\scriptsize H\one}$\tablenotemark{b} & 
$N_{\textup{\scriptsize SCC}}$ &	
\tmult{$S_N$}\\	

&				
\tmult{(kpc)} &		
(\Msun)     &			
(\Msun)     &			
\tmult{(\Msun\ /yr)} &	
\tmult{($\times10^{-11}\,$yr$^{-1}$)} &	
(\Msun\,yr$^-1\,$kpc$^{-2}$) & 
(\Msun\, pc$^{-2}$) & 
\tmult{~}& \\			
 \tableline
A	& 31	& 15	& 10.22~	& 9.0							& 3.88	& 0.47	& 23.46	& 4.25	& $-$2.3		& \phantom{$-$}0.1	& 50	& 0.4	& 0.2\\
B	& 20	& 13	& 9.95~	& $<8.1$\tablenotemark{c}\phantom{$<$}	& 0.23	& 0.02	& 2.56		& 0.42	& $-$3.1		& $-$0.4		& 28	& 2.5	& 0.9\tablenotemark{d}\\
C	& 29  	& 19	& 9.88~	& 9.5							& 2.06	& 0.17	& 27.33	& 4.34	& $-$2.5		& $-$2.7		& 133	& 0.3	& 0.2\\
D	& 16  	& 13	& 9.56~	& 8.9							& 0.43	& 0.04	& 11.89	& 1.90	& \phantom{$-$}0.7 &\phantom{$-$}0.6	& 45	& 2.5	& 0.9\tablenotemark{e}\\
\tableline
$\Sigma$  & 200 & 120 & 10.56~  &   9.7  &    6.60  &    0.50  &	&	&  &  &    256  &      \tmult{~}\\
 \tableline
\end{tabular}
\end{center}
$^{a,b}${The Schmidt-Kennicutt relation is tested on 
	spiral galaxies with inclination $i<75^\circ$; \galb\ 
	is a lenticular and \gala\ near the limiting inclination 
	examined by \citet{kennicutt98}.}\\
$^{c}${There is no detection of H\one\ in lenticular \galb\ 
	in our VLA data; this value represents an upper limit, 
	based on the sensitivity of our observations. }\\
$^{d,e}${The GC systems of \galb\ and \gald\ are 
	overlapping; this is the combined number.}

\end{table*}

\subsection{Galaxy B and the origin of group lenticulars}
The formation of lenticular galaxies in clusters is still somewhat puzzling, therefore interpreting the origin of \galb\ is of merit to understanding compact group evolution. In environments where major mergers are meant to be the norm, the evolution of a galaxy is likely to disrupt its disk structure -- although recent modeling works by \citet{combes08} and \citet{brooks09}  have found disk re-growth a possibility in mergers where one of the parties is gas-rich \citep[also discussed in the context of `the Mice', NGC~4676, by][]{barnes02}. {This process does, however, require a timescale of a few Gyr and would most likely leave the imprint of a bimodality in the GC population, which will be treated in \S~\ref{sec:gcs}}.

Validating such a scenario is beyond the reach of our diagnostic powers. Essentially, the galaxy shows no signs of massive cluster formation for at least the last $\sim3$~Gyr. This implies that it entered a quiescent phase at that time, perhaps signaling the depletion of fuel. As a simple exercise, we can estimate the amount of stellar mass the galaxy has produced since that point in time. Assuming a constant SFR over the last $3$~Gyr, equal to the rate currently observed, we derive a very low \mstar\ yield, suggesting that the galaxy produced $\sim$95\% of its photometrically inferred \mstar\ before that time. By observation, isolated spirals do not deplete their gas reservoir in a Hubble time, so \galb\ must have been aided in some way in processing the gas into stars. The presence of a disk structure {and dynamics} (\ie\ the distinguishing factor between ellipticals and lenticulars) implies that the current state was not reached through a major merger. 

In any case, the presence of an S0 type galaxy in a compact group that is apparently free of major interactions and mergers might suggest that lenticulars can indeed be sculpted by the conditions reigning in groups.

\section{The globular cluster system as a tracer of past star formation epochs}\label{sec:gcs}
The study of old globular clusters expands our window for galactic archaeology to a much earlier era. 
Our \hst\ imaging provides an exquisite opportunity to do so. In this section we provide a full account of the globular cluster system of HCG~7, constituting the first high resolution study of globulars in compact group spirals/lenticulars. 

\begin{figure*}
\begin{center}

	\includegraphics[width=0.32\textwidth, angle=0]{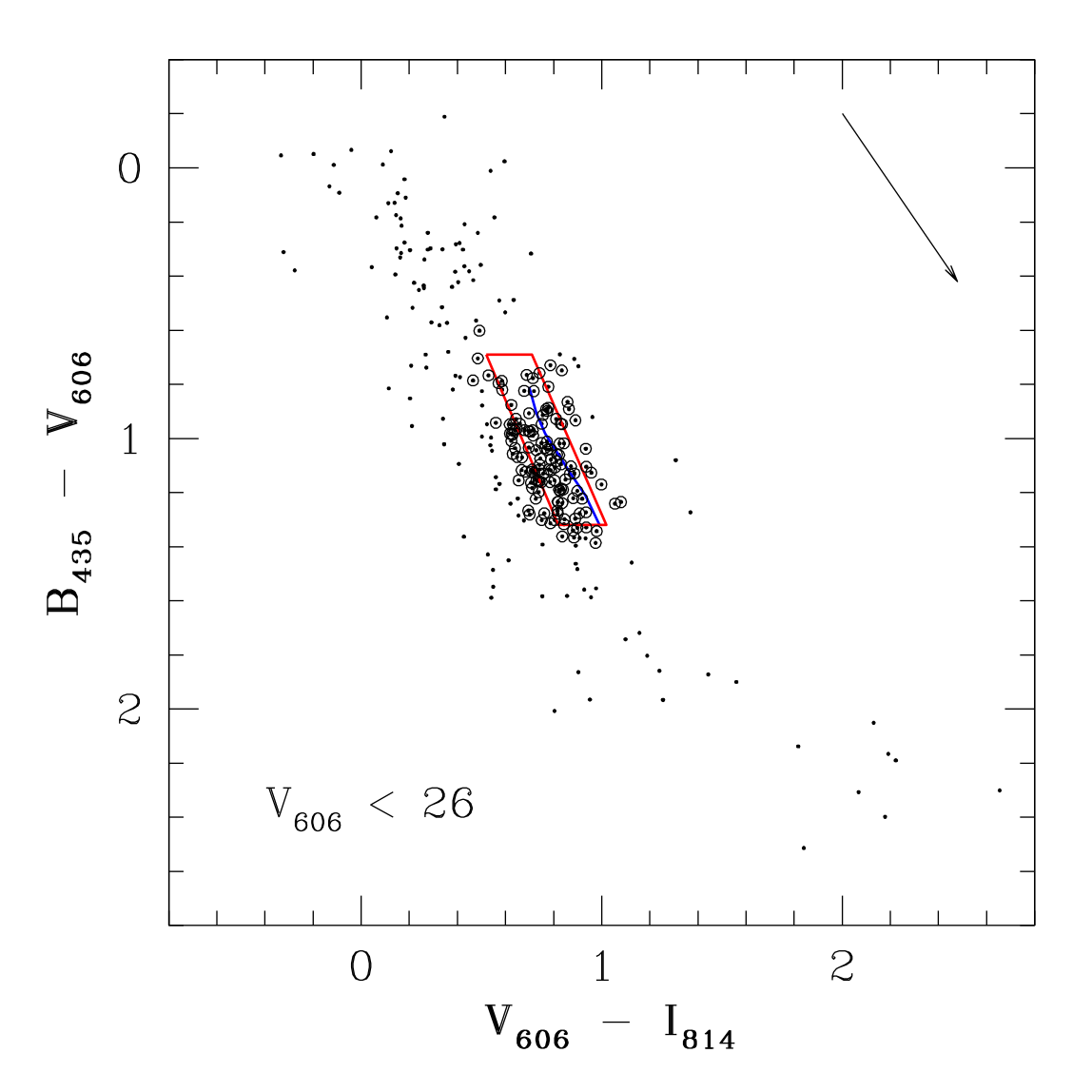}~
	\includegraphics[width=0.32\textwidth, angle=0]{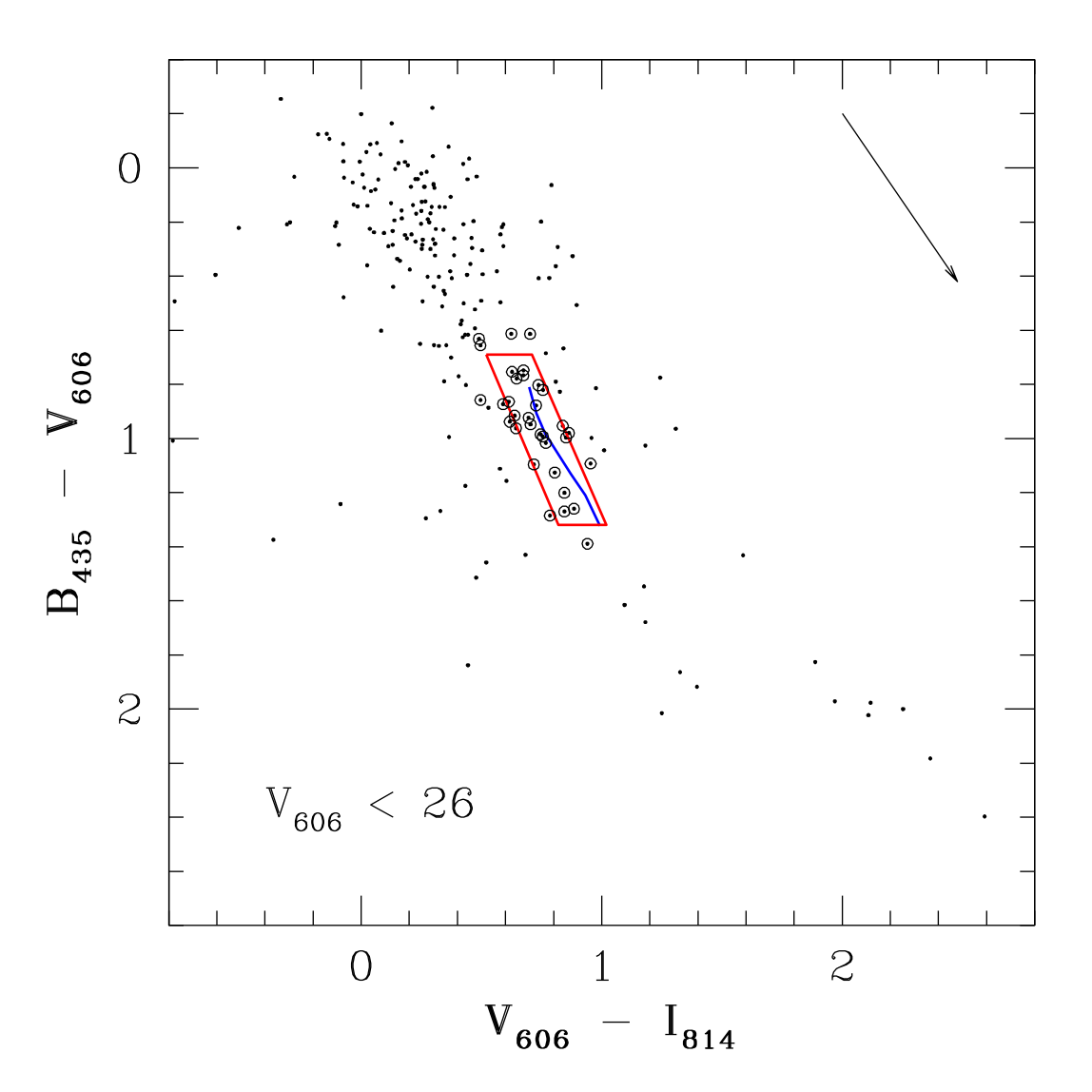}~
	\includegraphics[width=0.32\textwidth, angle=0]{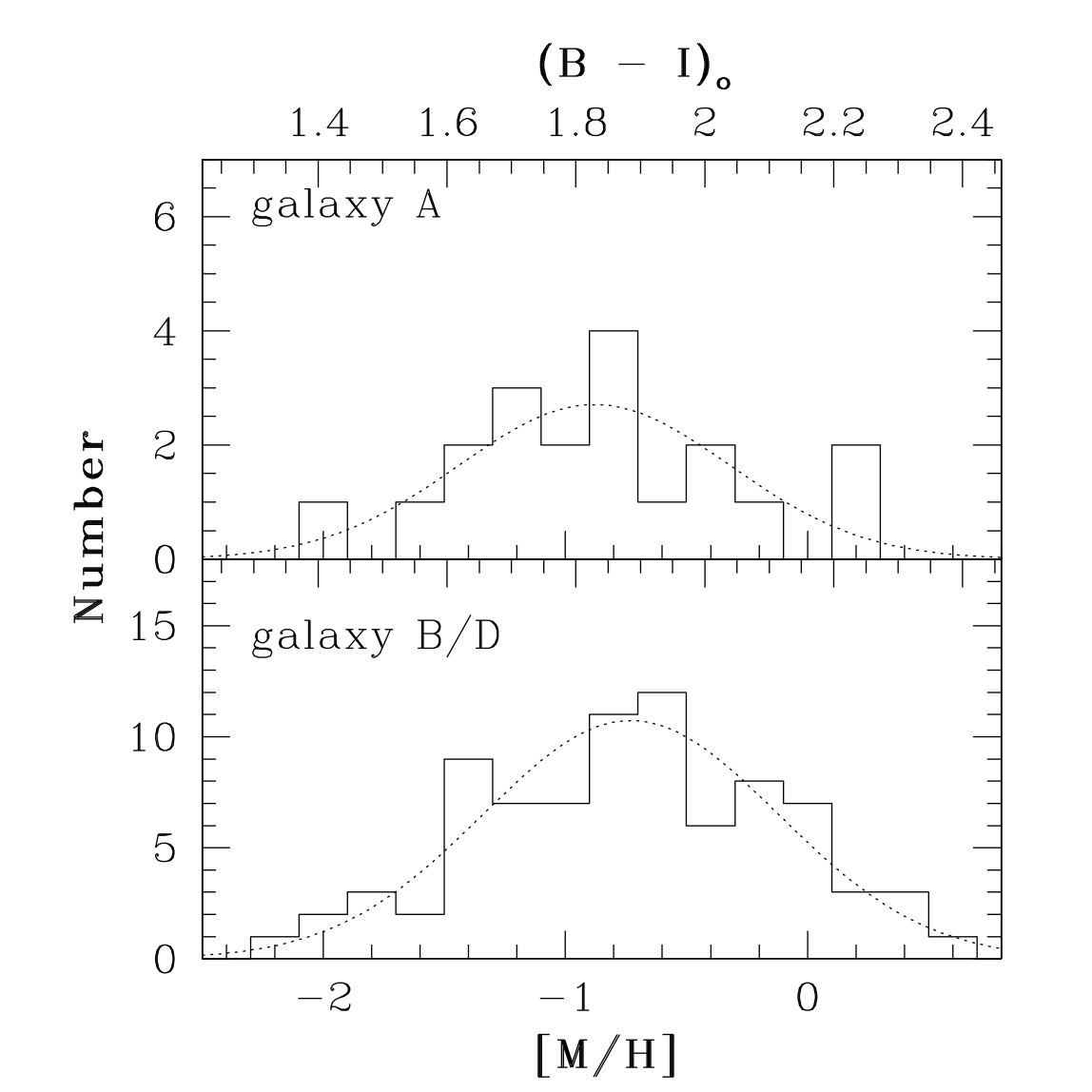}
	
	\caption{
	{\bf Left:}~Color-color (CC) diagrams for globular cluster candidates in fields one and two, containing \gala/\galb/\gald\ and \galc\ respectively. We have indicated the precise boundaries of the GC selection box. {The GC colors are consistent with Milky Way globulars}. 
	{\bf Right:}~The metallicity distribution of the \gala\ and \galb/\gald\ GC systems, as derived from the $B-I$ color. {We find no evidence of bimodality, a trait attributed to mergers in the early history of galaxies \citep{muratov10}}. This analysis could not be performed on the GC system of \galc, due to its face-on inclination (see Sec.~\ref{sec:gcs}). 
	}
	\label{fig:gcs}
	
\end{center}
\end{figure*}

The color-color diagrams of Figure~\ref{fig:gcs} show a significant population
of stellar objects with colors consistent with that of old ($\sim10-12$ Gyr) globular clusters; the majority are found around the lenticular \galb. We will treat the GC system of each galaxy separately, starting with the intertwined
population of GCs around \galb/D. 

\subsection{Galaxies B \& D}
From Fig.~\ref{fig:regions}, it is clear that the projected haloes of the 
SB0 galaxy \galb\ and the face-on spiral \gald\ overlap, making a clean separation 
of their respective GC systems difficult.  However, as \gald\ is a spiral 
that has a much lower luminosity than \galb~($M_V=-19.7$ \textit{vs.} $-21.0$ 
for \galb), the number of GC candidates expected in the halo of \gald\ 
should be quite small \footnote{
If we adopted $S_N=0.4$ for \gald\ 
alone \citep[\eg][]{goudfrooij03, chandar04}, we would expect 
to detect $\sim 13$ GC candidates in our ACS field.}.  
Thus we will 
associate the GC candidates surrounding galaxies \galb\ and \gald\ as being part of
\galb, after first excising 11 GC candidates that lie close to the
center of \gald\ -- these objects may be affected by variable
extinction internal to \gald.  Of the 93 GC candidates surrounding
\galb/\gald, 11 near \gald\ have been removed as noted above, leaving a
total of 82 objects.

As only a fraction of the GC system of \galb\ lies on our ACS
fields,we derive the expected total number of GCs around \galb\ by
counting the GCs in a wedge (of angular extent 113 degrees) that
extends to a distance of 30 kpc from the center of \galb, yet lies
wholly within the ACS field.  From the observed number of $71\pm 8$
objects that lie within this region, we predict a total of $227\pm 27$
objects (assuming radial symmetry).  Adopting a photometric
completeness fraction $f=0.9\pm0.1$ to $M_V<-7.9$ (see discussion above) and
correcting for the GCs below the magnitude cutoff, we derive
{a total number of} $630 \pm 230$ GCs in \galb.  For $M_V=-21.0$, this translates to
a specific frequency (the number of globular clusters per unit galaxy luminosity) of 
$S_N=2.5\pm 0.9$, in accord with global values of
$S_N$ in the literature for S0 galaxies \citep[both in the field and in the
cluster environment;][]{az98, kundu01}.
The values of $S_N$ for S0 galaxies are (in general) higher (by a
factor of $\sim 3$) than those for spiral galaxies, and this difference
has been attributed to the fading (after the depletion of the original
gas) of the stellar populations in spirals after a few Gyr \citep{aragon06, barr07} 
{and the consequent under-detection with respect to lenticulars}. 

Through the color distribution we are also able to derive the metallicity
distribution of the GCs in \galb.
Fig.~\ref{fig:gcs} shows 
the  $(B-I)_0$ colors of the candidates (transformed from the 
\bb~$-$~\ib\ colors using the synthetic transformations from Sirianni 
et al 2005).  The distribution does not show an obvious color bimodality 
common to many massive galaxies \citep[\eg][]{brodiestrader06, peng06}\footnote{
{We confirmed the unimodal nature of the distribution through} an attempt to to fit the color 
distribution with a dual-Guassian fit using the KMM algorithm 
\citep{ashman94}; we find that such a fit is not significantly 
better than a single-Gaussian fit ($p=0.65$).}. 
A single Gaussian fit yields a peak $(B-I)_0= 
1.88\pm 0.03$ (error in the mean value) and a dispersion of 0.23 mag.   
Under the assumption that all of the GC candidates have ages similar to 
that of Milky Way globular clusters ($\sim 12$~Gyr), we use the 
$(B-I)-$~[Fe/H] transformation from Harris~\etal~(2006):

\begin{displaymath}
	(B-I)_0 = 2.158 + 0.375~\textup{[Fe/H]}
\end{displaymath}
to convert these values to a peak [Fe/H]~$=-0.74\pm 0.08$ for the galaxy \galb\ 
GCs.    
This value is very similar to the peak metallicities of 
GCs in other E/S0 galaxies, both in the field and in the cluster 
environment \citep[\eg~][]{kundu01, peng06}. 

{Interestingly, recent modelling work by \citet{muratov10} interprets the bimodality of the GC color distribution to a few late mergers in the history of a given galaxy. If this is true, it lends strong support to our hypothesized lack of major mergers in HCG~7}. 

\subsection{Galaxy A}
The nearly edge-on \gala\ shows a small number ($21\pm 5$) 
of GC candidates, some projected onto its disk, others in the surrounding region. 
Adopting the same
methodology and corrections as outlined above, this translates to a
total expected number of $140\pm 70$ within its halo.
Assuming $M_V=-21.2$ gives rise to 
$S_N=0.4 \pm 0.2$.  To investigate the color distribution of the GC
system around \gala\, we first extracted a subsample of 19
candidates located outside the 
center of the galaxy, so as to limit the contamination from any
reddened younger clusters.  The resulting 
$(B-I)_0$
color histogram in 
Fig.~\ref{fig:gcs} 
shows a wide range of colors, indicative of a wide range of
metallicities.  The mean color of the GC candidates around \gala\ is
$(B-I)_0=1.83\pm 0.05$, {indicating} a mean metallicity of [Fe/H]~$=-0.88\pm 0.13$ 
\citep[assuming old cluster ages and the color-metallicity relation of][used 
above]{harris06}. 
While there is little more that can be gleaned from this 
small sample, it is worth noting that the wide range of metallicities 
of the GCs in \gala\ (the first time a metallicity distribution function has been constructed for 
GCs in a spiral in the compact group environment) is similar to that 
observed in other spiral galaxies of similar luminosity \citep[\eg][]{barmby2000, goudfrooij03, chandar04}.

\subsection{Galaxy C}
The orientation of face-on spiral \galc\ impedes the extraction of
GCs, as many clusters are expected to be heavily extinguished by
internal dust, or simply lost in the variable background.  Another
caveat is the potential sample contamination from heavily reddened
younger clusters; the lack of $U$ band information does not allow us
the ability to clearly distinguish these possibilities.  With
this in mind, the total number of point sources with GC-like colors
that lie within 30 kpc of \galc\ is 29.  Assuming the entire GC system
lies within our ACS fields, we derive a total GC population of
$N_T=80\pm 34$, \ie\ a specific frequency of $S_N=0.3\pm 0.2$. While
highly uncertain, this number is consistent with expectations for late-type 
spiral galaxies \citep[\eg][]{chandar04}.  As all of the GC
candidates are projected onto the disk of \galc\, we are unable to
extract a `clean' color/metallicity distribution.

\section{The multi-wavelegth view: spectra and SEDs}\label{sec:seds}
The spectral energy distribution (SED) of a galaxy provides valuable diagnostics of its overall properties. In this section we look at the emission originating from the four group galaxies in the optical and near/mid infrared ($0.4-24\,\mu$m). This is the combination of optical data from the SDSS, NIR (\textit{JHK}) data from 2MASS and mid-IR data from \spit\ (IRAC and MIPS; Section~\ref{sec:spit}). The near- and mid-IR data are drawn from \citet{johnson07}. We refer back to the optical spectra of galaxy nuclei to interpret the broadband emission and make use of the \airac\ SED diagnostic discussed in \citet{gallagher08} and to a lesser extent in Section~\ref{sec:spit}.  

\begin{figure*}
\begin{center}

	\includegraphics[width=0.7\textwidth, angle=0]{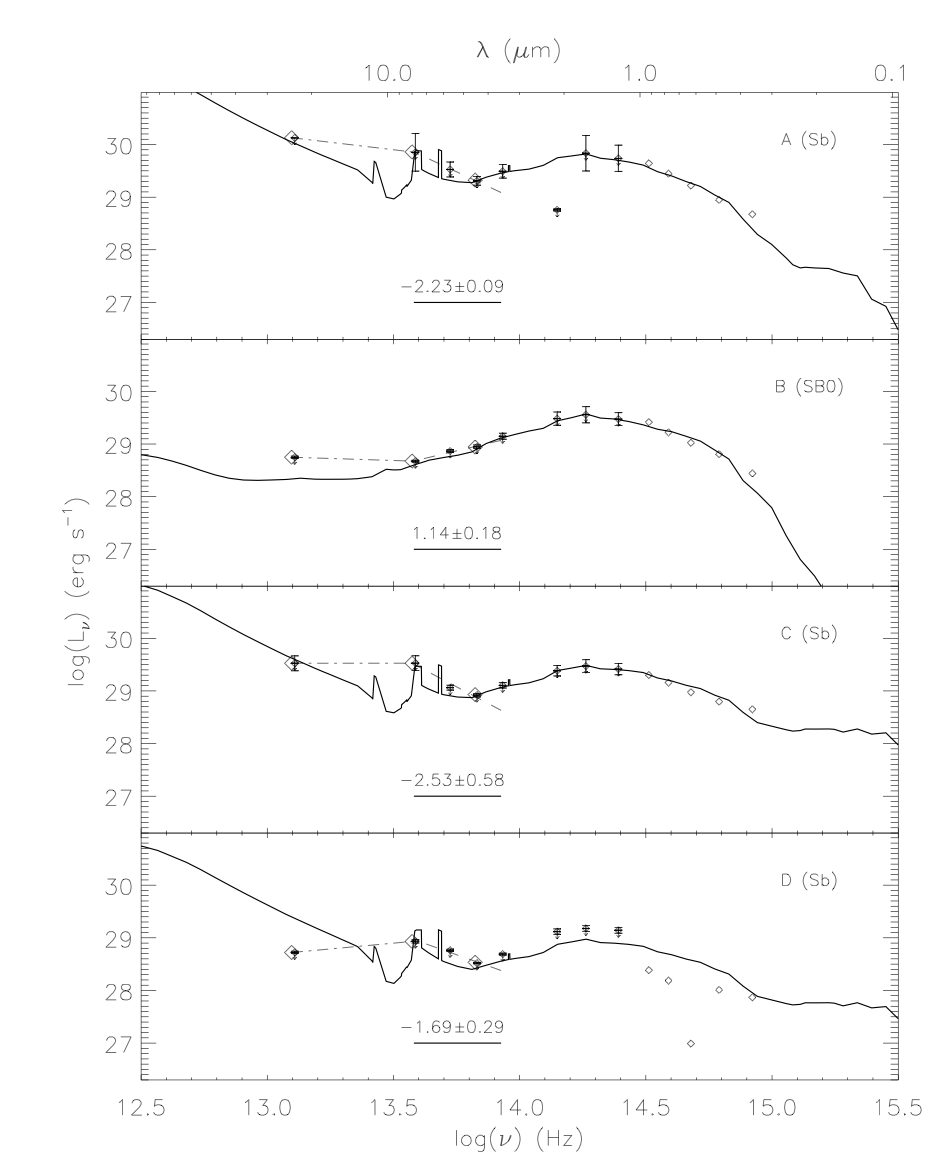}
	
	\caption{Optical to mid-IR SEDs for HCG~7 members. The photometric data, shown as diamonds, are drawn from the SDSS, 2MASS and \spit\ MIPS/IRAC. We annotate each panel with the adopted morphological type and with the value of the \airac\ diagnostic (the line under the \airac\ shows the fitting range). The lines represent GRASIL templates with parameters adapted to the derived characteristics of each galaxy (see Section~\ref{sec:seds}). In all, the template fits reveal different {spectral} types than {expected, based on morphology} for \galc, \gald\ and possibly also \gala. The lenticular \galb\ is found to be best described by a young elliptical spectrum, $\tau=8$~Gyr. The template plotted for \gala\ can be replicated by maintaining the original Sa designation and enhancing the star formation by a factor of $\sim5$. 	The SED of \gald\ is difficult to model, {perhaps due to its previous misclassification as SB0}. 
	}
	\label{fig:seds}
	
\end{center}
\end{figure*}

The SEDs, presented in Fig~\ref{fig:seds}, are fit with appropriately chosen optical-to-far-IR galaxy templates from the GRASIL library \citep{silva98}. They reveal some very interesting aspects of the galaxies, all of which show an excess or dearth of emission in one of the observed bands, when compared to `normal' templates -- \ie\ ones representative of each galaxy's morphological type (as given in \ned). To reconcile these discrepancies, we use appropriate `blends' of the individual emitting components in each GRASIL template: i) star-forming molecular clouds and the young stars contained therein; ii) light from evolved stars; iii) diffuse gas. The three components are referred to as `MC', `starlight' and `cirrus' in \citet{silva98}, the terminology that we hereforth adopt. 

In this Section we also provide a closer examination of the central spectra obtained with the Hydra multi-object spectrograph, described in Section~\ref{sec:hydra}. We also compare our results to those of a recent spectroscopic campaign to identify AGN in the full catalogue of HCGs \citep{martinez10}. 

\begin{figure*}
\begin{center}

	\includegraphics[width=0.7\textwidth, angle=0]{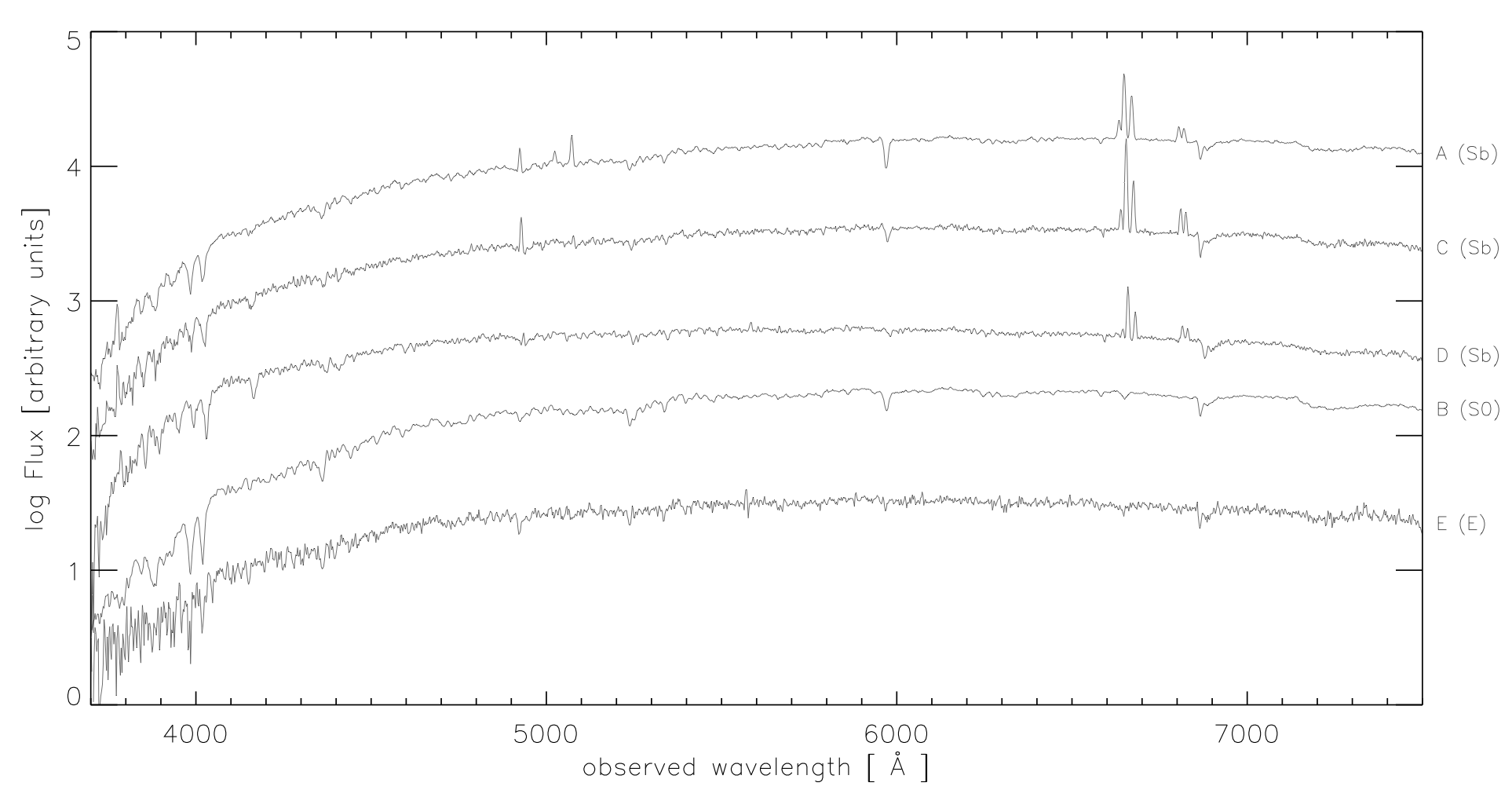}
	\includegraphics[width=0.7\textwidth, angle=0]{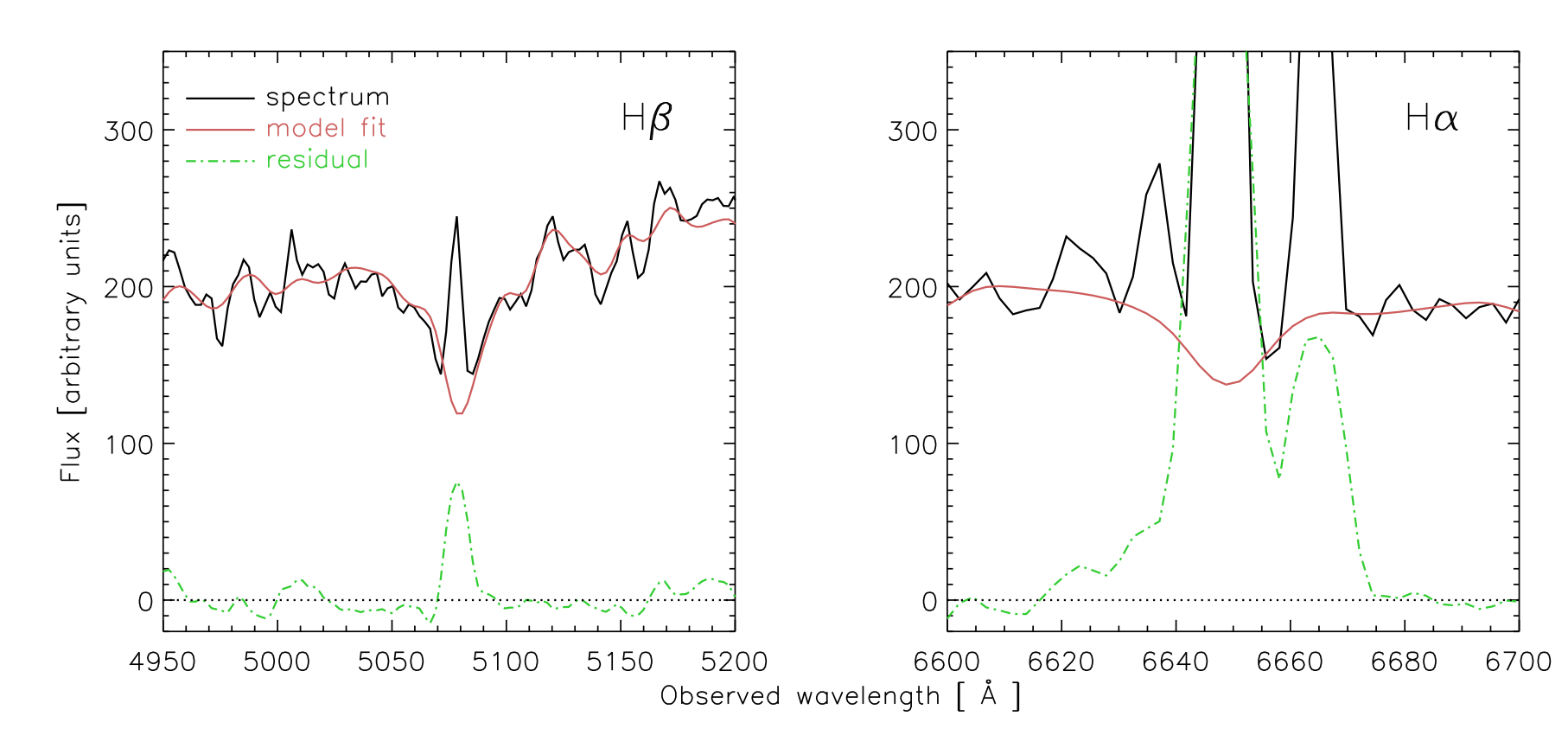}
	
	\caption{{\bf Top}~CTIO-Hydra optical spectra of the newly discovered compact galaxy \galx\ -- the only one of its kind detected down to our $R\sim20.5$ spectroscopic limit -- and the nuclei of the four main HCG~7 galaxies. The flux axis is plotted in logarithmic space to ease presentation and comparison. 
	The spectra of spiral galaxies \gala/\galc/\gald\ all exhibit \ha\ emission. The [O\three]/\hb, [N\two]/\ha\ and [S\two]/\ha\ emission line ratios of \gala\ place it in the H\two\ regime (see Section~\ref{sec:seds}), {indicating a high ionization continuum} in its core. \galc\ shows a spectrum more appropriate for an Sb galaxy, while \gald\ exhibits absorption in \hb, with a weak, embedded emission component, uncharacteristic of an Sc; we discuss the nature of this galaxy as a spiral at the verge of becoming a lenticular in Section~\ref{sec:seds}. The nucleus of \galb\ shows an elliptical-type spectrum, with deep absorption lines and no emission, as does the compact galaxy \galx, albeit at a significantly lower S/N$~\sim20$. The classification of all galaxies is discussed in Section~\ref{sec:seds} and Fig.~\ref{fig:seds}. 
	In the {\bf bottom} panels we demonstrate the `E+A' spectrum of \gald: the separate lines show the spectrum and pPxF model fits (see \S~\ref{sec:seds:d}) of the H$\beta$ and H$\alpha$ regions. 
	}\label{fig:spectra}
	
\end{center}
\end{figure*}

\subsection{Galaxy A}\label{sec:seds:a}
\gala\ is listed as type Sa in \ned; plotting the equivalent template produces a poor agreement with the observed mid-IR fluxes. Instead, we investigate the suitability of an Sb template. The 21-cm data reveal a low H\one\ content in the HCG~7 galaxies, {to the level of one third the nominal value \citep[][see \S~\ref{sec:gasmass}]{haynes84}}. Thus, we add only a third of the cirrus component to the template blend. Given the highly inclined line of sight towards the inner regions of \gala, we include only starlight extinguished by either dust or star forming, gas-rich regions (\ie\ no unextinguished starlight). Finally, we add the normal MC component. This template blend produces a very good match to the observed photometry. 

Before dismissing the Sa designation, however, we perform a test for a possible AGN in galaxy \gala. We measure the strengths of emission lines \hb, [O\three] $\lambda5007$, \ha, [N\two] $\lambda6584$ and [S\two] $\lambda\lambda6717,6731$. We then compare the log-ratios of [O\three]/\hb, [N\two]/\ha\ and [S\two]/\ha\ to the values in \citet[][we refer to their Fig.~4]{kewley06} to identify the presence of an AGN. Both the $\log$~([N\two]/\ha) and $\log$~([O\three]/H$\beta$) fall in the interface between H\two\ nuclei and AGN. The likelihood of \gala\ hosting an AGN is lessened by the value of the $\log$~([S\two]/\ha), that places it clearly in the area occupied by H\two\ nuclei. We can therefore attribute the high star formation activity in \gala\ to a central starburst. The \citet{martinez10} study classifies \gala\ as a `transition object' (TO), one that hosts an AGN and a circumnuclear starburst. {This scenario is consistent with the mid-IR morphology of the galaxy disk, which shows bright central emission across the spectrum and $24\,\mu$m emission in a limited halo. We also see a circumnuclear ring of stellar photospheric ($3.5\,\mu$m) emission (also detected as a faint region in the UV), which we interpret as low or even null star formation. This region is also devoid of gas down to our detection limit, indicating that star formation has slowed down or ceased around the nucleus either because the gas was spent in making stars, or that it was swept inwards, perhaps owing to the presence of the bar}. 

With this information in mind, we construct a GRASIL Sa template with an MC (star formation) component  boosted by five times the nominal Sa rate. In order to account for the excess dust produced in a starburst, we only include extinguished starlight. Finally, we maintain the cirrus emission to a third of the template value. This recipe {does not provide a better fit to the data and is equivalent to the Sb template} plotted in Fig.~\ref{fig:seds}. However, while the high inclination does not allow us to discern between these two distinct scenarios, {the disk morphology does support currently strong central activity}.

\subsection{Galaxy B} 
\galb\ matches the GRASIL 13~Gyr-old elliptical galaxy template extremely well, except in the $24\,\mu$m band, where it shows excess emission. In order to reconcile this effect, we compare it to a younger model, a distinct possibility for this lenticular galaxy. The presence of old globulars in this galaxy (Section~\ref{sec:gcs}) marks star formation activity at least $\gtrsim5$~Gyr ago. Beyond this point, the evolution of GC \textit{BVI} colors is very slow, therefore we can not certainly distinguish the ages of  GCs between $5-13$~Gyr (as discussed above). The best agreement in the $24\,\mu$m emission is offered by an  8~Gyr old elliptical template, although an age of 5~Gyr is not ruled out based on the SED.

\subsection{Galaxy C} 
\galc\ and \gald\ both display a dearth of $24\,\mu$m emission. The spectrum of \galc\ is not dominated by blue continuum, as one would expect for the nucleus of an Sc galaxy. Instead, the overall shape is more reminiscent of an Sb-type. There is booming \ha\ emission, {with an equivalent width of $-29$~\AA}, however the emission-line diagnostics do not show evidence of {an intense ionization continuum}. M10 classify the nucleus of \galb\ as star forming, in tune with our analysis.

The reduced $24\,\mu$m emission is reconciled by adopting an Sb-type template with low cirrus emission; this is the one shown in Fig~\ref{fig:seds}.

\subsection{Galaxy D}\label{sec:seds:d}
This galaxy presents a very interesting `E+A' central spectrum. Adopting for the moment its \ned\ designation as Sc, one would expect to see hydrogen in strong emission lines, a flat spectral shape and plenty of emission features below $\lambda3700$. Instead, the spectrum of \gald\ features moderate \ha\ emission and \hb\ mostly in absorption, with only a weak emission feature. The upper series Balmer lines are seen in pure absorption, all the way to H$\epsilon$, evident to the side of a weak Ca\two\ K interstellar absorption line. The dominance of the stellar H$\epsilon$ line over Ca\two\ H suggests a low ISM content of the region covered in the spectrum, if not the entire galaxy. {This is reflected in the reduced $24\,\mu$m emission, which also originates} from interstellar material, namely medium-size dust grains {in the case of $24\,\mu$m}. 

`E+A' spectra normally arise from a {relatively recent and} abrupt cessation of star formation activity in a previously {star-forming} galaxy. {What little star formation persists produces the mission component, which becomes embedded in the deep absorption lines of the recently formed stellar population and the underlying old stars}. 

{We investigate the spectrum in this light through a model fit. We used the `penalized pixel fitting method' \citep{ppxf} to fit around the emission component and construct a template of the underlying continuum and absorption lines, as shown in the bottom panels of Fig.~\ref{fig:spectra}. We then subtracted this template from the observed spectrum and measured the equivalent widths (EW) of the H$\delta$ in absorption and H$\alpha$ in emission as $5.23\pm0.22$~\AA\ and $-5.58\pm0.03$~\AA\ respectively. These lines are not as wide as the archetypal post-burst galaxies selected by \citet{zabludoff96}, implying that the recent epoch of star-forming activity was not quite as strong as a starburst. }

Given {the above, as well as the} low gas and ISM content of the galaxy and its weak, {asymmetric} spiral structure, we suggest that this system is in a transitory stage between spiral and lenticular. At the current SFR this transition would take $\log\tau\sim9.2-9.4~$yr (Section~\ref{sec:gasmass}); this is an upper limit, as another interaction-triggered starburst would shorten this timescale. 

The disagreement between template and observation in the optical regime is due to the previous misclassification of this galaxy as SB0: SDSS photometry is conducted in a manner consistent with morphological type, meaning that any emission from the (faint) spiral arms of this galaxy would have been omitted.

\section{Implications for the evolution of HCG~7}\label{sec:discussion}
\subsection{The effect of environment on the evolution of HCG~7}
In Section~\ref{sec:gasmass} we discussed the H\one\ distribution about the group. {The VLA data do not detect an IGM in neutral gas} and thus suggest that the HCG~7 galaxies have not undergone any major interactions recently, as such intense dynamical processes leave evidence for a few Gyr \textit{post facto}. This was shown in the local universe by \citet[][part of the PAndAS project]{mcconnachie09} who recently presented the {purely} stellar tidal streams of M31 as the direct result of a $\sim3$~Gyr old encounter with M33. Naturally, with sources at distances of $\lesssim1$~Mpc, the  PAndAS survey reaches much fainter detection limits compared to this study. Gaseous tidal features should, however, be detectable $\sim1$~Gyr after the interaction in H\one, as discussed in Section~\ref{sec:gasmass}. The detection limit of the VLA data we use in this study is lower than the $10^{20}~$cm$^{-2}$ level reached in a Jodrell Bank survey of the M81 group \citep{boyce01}, which uncovered previously undetected tidal features. {In addition, the} \textit{Rogues Gallery} \citep{rogues}, showed very accentuated features at the $3\times10^{19}~$cm$^{-2}$ level ($\simeq0.8~$\Msun\,pc$^{-2}$), {a level easily accessible by our survey}.

One additional caveat is the survivability of tidal features in relation to the environment: \citep[\eg][]{mihos01} showed that their lifetime is curtailed in crowded environments. {The \hst\ data and the wide-field imaging we analyzed in Section~\ref{sec:bplusr} address this issue:  high SFR epochs are marked by massive clusters, the largest and brightest of which register as point sources}. With the ability to {confidently distinguish star clusters} down to the 24~mag level ($M_B=-9$~mag), we can exclude major recent episodes of star formation {in tidal debris}. 

The \chan\ images treated by Tzanavaris~\etal~(in preparation) show the four galaxies, but no IGM. 
This null detection of an X-ray IGM indicates that the missing H\one\ is
not concealed in a hot plasma phase that has been stripped from the
galaxies during previous interactions. 

At the same time, the current and past starbusts in \gala/\gald\ are telling of active star formation activity across the group. In the absence of major interactions, AGN, a hot IGM and any other obvious energizing impulse, internal or external, we resort to the complex tidal field for an explanation of this activity. Simulations by \citet{martig08} find the intricate tidal field of groups and clusters to enhance bursts of merger-induced star formation by a factor of up to 6, although the exact physical mechanism behind this is not well understood. It might therefore be safe to assume a similar enhancement of star formation due to minor interactions between galaxies, {since galaxies undergoing mergers and interactions are subject to similar gravitational processes; the result might not be the same, but the mechanism is similar}. In this scenario, a grazing encounter can enhance the star formation rate {compared to an identical system in isolation} without the stripping and redistribution of gaseous material. 

This is in accord with the study by \citet{walker10} who find evidence for  
rapid evolution in the compact group environment {compared to clusters and the field}. The mechanism is the depletion of the gas reservoir in a manner far quicker than field galaxies. However, like in the models by \citet{martig08}, it is extremely difficult to pinpoint the physical process behind this relatively active star formation.

\textit{Ergo}, we can conclude that this system is evolving quickly, under the influence of the tidal field; our observations suggest that a recent minor interactions have given rise to {enhanced star formation, currently in \gala, and possibly recently in \gald}. 

\subsection{The fossil group: on the path toward a dry merger?}
The information derived in this work constrains the state of HCG~7 as a moderately star forming, H\one-deficient system. Whatever the mechanism driving its evolution, by combining the derived measurements we can make a prediction of the group's future. Considering the available reservoir of H\one\ and H$_2$, assuming that the galaxies are free to sustain their current rate of star formation (an upper limit in the case of \gala), they will convert the gas to stars within $\sim2$~Gyr. This timescale, comparable to the crossing time ({derived as $R/\sigma\lesssim2$~Gyr in \S~\ref{sec:gasmass}}), may be too short for a series of major interactions to occur. 
	
	Following this hypothetical scenario, when the galaxies eventually converge to their common barycenter they would be devoid of gas and will therefore have no further material eligible for star formation. 
	Such conditions would result in a `dry merger', whereby there is virtually no gas involved in the interaction \citep[\eg][]{vandokkum05}. This would give rise to an elliptical galaxy, as per the regular end-state scenario for a galaxy group, however, without a bright X-ray halo, a feature that normally arises through the processing of intra-group gas. {It must be stressed here, however, that the IGM is most likely present in compact groups, but due to their shallow potential, it was never heated to emitting temperatures and is not detectable. This means that a major fraction of the X-ray halo of the end-state of a compact group will arise from this material. In this sense, the following consideration only applies to the evolution of compact groups, or arrangements of galaxies with a similar velocity dispersion. }
	
	We can test the validity of this hypothesis in terms of observations and theoretical predictions of the X-ray-to-blue luminosity ratio, $L_X/L_B$. Continuing on the dry merger toy model, let us assume that all gas will be internally processed, \ie\ converted into stars, ahead of the final merger. In this case, the galaxy would have faint X-ray emission originating from its stellar halo. Given the stellar and gas masses in Table~\ref{tab2}, we derive a total mass of $\log M_\textup{\scriptsize TOT}=10.70$, a medium-sized elliptical, considering that they reach masses in the order of $10^{13}$~\Msun. This value accounts for the contribution to the total mass by helium through the relation $M_\textup{\scriptsize He$+$H\one} = 1.4\times$\mhi\ \citep{mcgaugh00} . We can now estimate the blue luminosity through the mass-to-light ratio of a BC03 model SSP of age 12~Gyr\footnote{This is clearly an oversimplification, as the hypothesized population would be a blend of stars of different ages. However, the small amount of gas left in the system at the present time would only give rise to a small young star population, thus the light would be dominated by old stars}: \mstar$/L_B\simeq2.0$, therefore $\log L_B=10.07$. Observations by \citet{osullivan07} provide a range of corresponding $\log L_X=39.5-41.5$~erg\,sec$^{-1}$. Models by \citet{cox06} of mergers between disk galaxies result in the middle of that range, $\log L_X\simeq41$~erg\,sec$^{-1}$. A more conservative approach, where half the gas is used up in star formation and the rest released in the IGM through interactions leads to \mstar$=10.64$, $\log L_B=10.35$ and $\log L_X\simeq40.5$~erg\,sec$^{-1}$ \citep[from][]{cox06} and a weak X-ray IGM signature. Note that this emission is only due to the stellar X-ray halo, not the gas in the IGM. In either scenario, the X-ray component is bound to be weak, because of the low mass of the fossil group. This is discussed in more detail in \citet{gallagher10} for the case of HCG~31. 
	
	In all, {both scenarios} are in excellent agreement with the two-pronged evolutionary sequence we presented in Section~\ref{sec:seq} and Fig.~\ref{fig:sequence}, as well as the sequence proposed by \citet{verdes01}. In that arrangement, HCG~7 lies between Hickson groups 2 and 90: the former a collection of undisturbed, star-forming galaxies, the latter an apparent ongoing merger between two lenticular galaxies and a dusty spiral. It is important to note that this is a qualitative evolutionary sequence, as the masses of HCGs~2 and 90 are not in agreement with that of HCG~7. The arrangement of galaxies in the schematic of Fig.~\ref{fig:sequence} argues that the distribution of gas {and arrangement of galaxies} is crucial in defining the end-state of a galaxy group. 
	
The need for dry mergers in the history of the universe is exemplified by the large scatter of observed X-ray luminosities in truly isolated ellipticals \citep[\cf\ recent observations presented by][]{osullivan07,memola09}. {They are also important in populating the `red sequence' \citep{bell04} and shaping the galaxy luminosity function}. The possibility of systems like HCG~7 evolving in the way we suggest here would thus present an opportunity to `jump' back to $z\sim1-2$ and observe {X-ray weak} ellipticals in the making.

\section{Summary}\label{sec:summary}
{We have} presented a multi-wavelength study of Hickson Compact Group~7. At first glance, this close grouping of four disk galaxies (three spirals, one lenticular) presents an overall seemingly undisturbed morphology: the individual members exhibit fairly regular disk/spiral structures and there are no evident tidal features in the optical, infrared, or radio data. Adding to this, the X-ray map of the group reveals no hot intra-group medium. Finally, the age distributions of star cluster populations in the individual galaxies show no over-densities and hence no eras of pronouncedly elevated star-formation rates. At this point, one might conclude that HCG~7 consists of four neighbors that bizarrely remain `strangers', despite having spent billions of years in a volume that can hardly accommodate them comfortably. 

Closer inspection reveals a different picture: spectra of the central regions of the galaxies uncover {enhanced star formation activity} in the past and present; at its lowest level, optical $B+R$ imaging shows faint stellar streamers connecting \gald\ and the newly confirmed compact member \galx\ to the evolved \galb; and the H\one\ gas is not only heavily depleted -- as confirmed by the UV-to-IR SEDs of all galaxies -- but also distributed in truncated disks that fall short of the galaxies' optical radii. {In addition, we find evidence of} fast gas depletion (compared to secularly evolving field galaxies), {another indicator of past interactions}, yet no trace of an H\one\ or X-ray intragroup medium. Our diagnostic tools are powerful enough to detect such features, therefore we conclude they are not there to be found. {Finally, the unimodal color distribution of the GC population in galaxy \galb\ might suggest a lack of major mergers in its history, according to the cosmological scale models of GC formation and evolution by \citep{muratov10}}. This begs the question of why the galaxies are evolving in a manner similar to closely interacting systems. 

While the physical mechanism behind this behavior is ultimately unclear, it is undoubtedly linked to the compact group environment. This is in tune with recent models of merger-induced star formation in compact groups, where the star formation rate can be boosted by a factor of $\sim6$, {as compared to a merger occurring in the field} \citep{martig08}. 

We conclude this study by proposing a toy model for the future evolution and therefore the end-state of this compact group. Our observations imply short depletion timescales for the little remaining gas. Assuming that the galaxies will not undergo major interactions or mergers in the next crossing time -- as they have not during the last $\sim\tau_\textup{\scriptsize cross}$ -- there will be little gas in the eventual merger between the four galaxies. This paints the picture of the fossil group as an elliptical of average mass and with a faint (or no) X-ray component, originating solely from its stellar halo. This would differentiate it from the more {studied} evolutionary path that results in a gas-rich merger and a bright X-ray halo that arises chiefly from IGM gas.

This image is in tune with the position of HCG~7 in the top prong of our proposed evolutionary sequence of compact groups. HCG~90, a representative end-state for this sequence appears to be coming together with no detectable X-ray and H\one\ components. 
This prospect may be extremely important in understanding the X-ray-to-optical luminosity ratio, $L_X/L_B$, observed at $z=0$. 

\section*{Acknowledgements}
	ISK would like to thank Amanda Kepley for help with the interpretation of the radio data. 
	We would also like to thank Rodrigo Carrasco for useful discussions regarding the use of the Hydra \texttt{iraf} package. 
	Support for this work was provided by NASA through grant number HST-GO-10787.15-A from the Space Telescope Science Institute which is operated by AURA, Inc., under NASA contract NAS 5-26555, and the National Science and Engineering Council of Canada (SCG \& KF). 
	PRD would like to acknowledge support from \hst\ grant HST-GO-10787.07-A.
	KEJ gratefully acknowledges support for this work provided by NSF through CAREER award 0548103 and the David and Lucile Packard Foundation through a Packard Fellowship.


\bibliographystyle{apj}
\bibliography{references}

\begin{thebibliography}{118}
\expandafter\ifx\csname natexlab\endcsname\relax\def\natexlab#1{#1}\fi

\bibitem[{{Anders} {et~al.}(2004){Anders}, {de Grijs}, {Fritze-v.~Alvensleben},
  \& {Bissantz}}]{anders04b}
{Anders}, P., {de Grijs}, R., {Fritze-v.~Alvensleben}, U., \& {Bissantz}, N.
  2004, MNRAS, 347, 17

\bibitem[{{Arag{\'o}n-Salamanca} {et~al.}(2006){Arag{\'o}n-Salamanca},
  {Bedregal}, \& {Merrifield}}]{aragon06}
{Arag{\'o}n-Salamanca}, A., {Bedregal}, A.~G., \& {Merrifield}, M.~R. 2006,
  A\&A, 458, 101

\bibitem[{{Ashman} {et~al.}(1994){Ashman}, {Bird}, \& {Zepf}}]{ashman94}
{Ashman}, K.~M., {Bird}, C.~M., \& {Zepf}, S.~E. 1994, AJ, 108, 2348

\bibitem[{{Ashman} \& {Zepf}(1998)}]{az98}
{Ashman}, K.~M., \& {Zepf}, S.~E. 1998, {Globular Cluster Systems}

\bibitem[{{Barmby} {et~al.}(2000){Barmby}, {Huchra}, {Brodie}, {Forbes},
  {Schroder}, \& {Grillmair}}]{barmby2000}
{Barmby}, P., {Huchra}, J.~P., {Brodie}, J.~P., {Forbes}, D.~A., {Schroder},
  L.~L., \& {Grillmair}, C.~J. 2000, AJ, 119, 727

\bibitem[{{Barmby} {et~al.}(2006){Barmby}, {Kuntz}, {Huchra}, \&
  {Brodie}}]{barmby06}
{Barmby}, P., {Kuntz}, K.~D., {Huchra}, J.~P., \& {Brodie}, J.~P. 2006, AJ,
  132, 883

\bibitem[{{Barnes}(2002)}]{barnes02}
{Barnes}, J.~E. 2002, MNRAS, 333, 481

\bibitem[{{Barr} {et~al.}(2007){Barr}, {Bedregal}, {Arag{\'o}n-Salamanca},
  {Merrifield}, \& {Bamford}}]{barr07}
{Barr}, J.~M., {Bedregal}, A.~G., {Arag{\'o}n-Salamanca}, A., {Merrifield},
  M.~R., \& {Bamford}, S.~P. 2007, A\&A, 470, 173

\bibitem[{{Bastian} {et~al.}(2005){Bastian}, {Gieles}, {Efremov}, \&
  {Lamers}}]{bastian05b}
{Bastian}, N., {Gieles}, M., {Efremov}, Y.~N., \& {Lamers}, H.~J.~G.~L.~M.
  2005, A\&A, 443, 79

\bibitem[{{Bastian} {et~al.}(2009){Bastian}, {Trancho}, {Konstantopoulos}, \&
  {Miller}}]{bastian09antennae}
{Bastian}, N., {Trancho}, G., {Konstantopoulos}, I.~S., \& {Miller}, B.~W.
  2009, ApJ, 701, 607

\bibitem[{{Bell} {et~al.}(2004){Bell}, {Wolf}, {Meisenheimer}, {Rix}, {Borch},
  {Dye}, {Kleinheinrich}, {Wisotzki}, \& {McIntosh}}]{bell04}
{Bell}, E.~F., {Wolf}, C., {Meisenheimer}, K., {Rix}, H., {Borch}, A., {Dye},
  S., {Kleinheinrich}, M., {Wisotzki}, L., \& {McIntosh}, D.~H. 2004, ApJ, 608,
  752

\bibitem[{{Bergvall} {et~al.}(2003){Bergvall}, {Laurikainen}, \&
  {Aalto}}]{bergvall03}
{Bergvall}, N., {Laurikainen}, E., \& {Aalto}, S. 2003, A\&A, 405, 31

\bibitem[{{Bigiel} {et~al.}(2008){Bigiel}, {Leroy}, {Walter}, {Brinks}, {de
  Blok}, {Madore}, \& {Thornley}}]{bigiel08}
{Bigiel}, F., {Leroy}, A., {Walter}, F., {Brinks}, E., {de Blok}, W.~J.~G.,
  {Madore}, B., \& {Thornley}, M.~D. 2008, AJ, 136, 2846

\bibitem[{{Binggeli} {et~al.}(1987){Binggeli}, {Tammann}, \&
  {Sandage}}]{binggeli87}
{Binggeli}, B., {Tammann}, G.~A., \& {Sandage}, A. 1987, AJ, 94, 251

\bibitem[{{Blanc} {et~al.}(2009){Blanc}, {Heiderman}, {Gebhardt}, {Evans}, \&
  {Adams}}]{blanc09}
{Blanc}, G.~A., {Heiderman}, A., {Gebhardt}, K., {Evans}, N.~J., \& {Adams}, J.
  2009, ApJ, 704, 842

\bibitem[{{Borthakur} {et~al.}(2009){Borthakur}, {Yun}, \&
  {Verdes-Montenegro}}]{borthakur09}
{Borthakur}, S., {Yun}, M.~S., \& {Verdes-Montenegro}, L. 2009, ArXiv e-prints

\bibitem[{{Boyce} {et~al.}(2001){Boyce}, {Minchin}, {Kilborn}, {Disney},
  {Lang}, {Jordan}, {Grossi}, {Lyne}, {Cohen}, {Morison}, \&
  {Phillipps}}]{boyce01}
{Boyce}, P.~J., {Minchin}, R.~F., {Kilborn}, V.~A., {Disney}, M.~J., {Lang},
  R.~H., {Jordan}, C.~A., {Grossi}, M., {Lyne}, A.~G., {Cohen}, R.~J.,
  {Morison}, I.~M., \& {Phillipps}, S. 2001, ApJL, 560, L127

\bibitem[{{Brodie} \& {Strader}(2006)}]{brodiestrader06}
{Brodie}, J.~P., \& {Strader}, J. 2006, ARA\&A, 44, 193

\bibitem[{{Brooks} {et~al.}(2009){Brooks}, {Governato}, {Quinn}, {Brook}, \&
  {Wadsley}}]{brooks09}
{Brooks}, A.~M., {Governato}, F., {Quinn}, T., {Brook}, C.~B., \& {Wadsley}, J.
  2009, ApJ, 694, 396

\bibitem[{{Bruzual} \& {Charlot}(2003)}]{bc03}
{Bruzual}, G., \& {Charlot}, S. 2003, MNRAS, 344, 1000

\bibitem[{{Cappellari} \& {Emsellem}(2004)}]{ppxf}
{Cappellari}, M., \& {Emsellem}, E. 2004, PASP, 116, 138

\bibitem[{{Chandar} {et~al.}(2004){Chandar}, {Whitmore}, \& {Lee}}]{chandar04}
{Chandar}, R., {Whitmore}, B., \& {Lee}, M.~G. 2004, ApJ, 611, 220

\bibitem[{{Combes}(2008)}]{combes08}
{Combes}, F. 2008, ArXiv e-prints

\bibitem[{Cox {et~al.}(2006)Cox, Matteo, Hernquist, Hopkins, Robertson, , \&
  Springel}]{cox06}
Cox, T.~J., Matteo, T.~D., Hernquist, L., Hopkins, P.~F., Robertson, B., , \&
  Springel, V. 2006, The Astrophysical Journal, 643, 692

\bibitem[{{de Vaucouleurs} {et~al.}(1991){de Vaucouleurs}, {de Vaucouleurs},
  {Corwin}, {Buta}, {Paturel}, \& {Fouque}}]{rc3}
{de Vaucouleurs}, G., {de Vaucouleurs}, A., {Corwin}, Jr., H.~G., {Buta},
  R.~J., {Paturel}, G., \& {Fouque}, P. 1991, {Third Reference Catalogue of
  Bright Galaxies} (Volume 1-3, XII, 2069 pp.~7 figs..~ Springer-Verlag Berlin
  Heidelberg New York)

\bibitem[{{Efremov} {et~al.}(1986){Efremov}, {Ivanov}, \&
  {Nikolov}}]{efremov86}
{Efremov}, Y.~N., {Ivanov}, G.~R., \& {Nikolov}, N.~S. 1986, in IAU Symposium,
  Vol. 116, Luminous Stars and Associations in Galaxies, ed. C.~W.~H. {de
  Loore}, A.~J. {Willis}, \& P.~{Laskarides}, 389--+

\bibitem[{{Eke} {et~al.}(2004){Eke}, {Baugh}, {Cole}, {Frenk}, {Norberg},
  {Peacock}, {Baldry}, {Bland-Hawthorn}, {Bridges}, {Cannon}, {Colless},
  {Collins}, {Couch}, {Dalton}, {de Propris}, {Driver}, {Efstathiou}, {Ellis},
  {Glazebrook}, {Jackson}, {Lahav}, {Lewis}, {Lumsden}, {Maddox}, {Madgwick},
  {Peterson}, {Sutherland}, \& {Taylor}}]{eke04}
{Eke}, V.~R., {Baugh}, C.~M., {Cole}, S., {Frenk}, C.~S., {Norberg}, P.,
  {Peacock}, J.~A., {Baldry}, I.~K., {Bland-Hawthorn}, J., {Bridges}, T.,
  {Cannon}, R., {Colless}, M., {Collins}, C., {Couch}, W., {Dalton}, G., {de
  Propris}, R., {Driver}, S.~P., {Efstathiou}, G., {Ellis}, R.~S.,
  {Glazebrook}, K., {Jackson}, C., {Lahav}, O., {Lewis}, I., {Lumsden}, S.,
  {Maddox}, S., {Madgwick}, D., {Peterson}, B.~A., {Sutherland}, W., \&
  {Taylor}, K. 2004, MNRAS, 348, 866

\bibitem[{{Ellison} {et~al.}(2009){Ellison}, {Simard}, {Cowan}, {Baldry},
  {Patton}, \& {McConnachie}}]{ellison09}
{Ellison}, S.~L., {Simard}, L., {Cowan}, N.~B., {Baldry}, I.~K., {Patton},
  D.~R., \& {McConnachie}, A.~W. 2009, MNRAS, 396, 1257

\bibitem[{{Elmegreen} \& {Elmegreen}(1983)}]{elmegreen83}
{Elmegreen}, B.~G., \& {Elmegreen}, D.~M. 1983, MNRAS, 203, 31

\bibitem[{{Elmegreen} \& {Salzer}(1999)}]{elmegreen99}
{Elmegreen}, D.~M., \& {Salzer}, J.~J. 1999, AJ, 117, 764

\bibitem[{{Fruchter} \& {Sosey}(2009)}]{multidrizzle}
{Fruchter}, A., \& {Sosey}, M. 2009, {The MultiDrizzle Handbook}

\bibitem[{{Gallagher} {et~al.}(2001){Gallagher}, {Charlton}, {Hunsberger},
  {Zaritsky}, \& {Whitmore}}]{gall01}
{Gallagher}, S.~C., {Charlton}, J.~C., {Hunsberger}, S.~D., {Zaritsky}, D., \&
  {Whitmore}, B.~C. 2001, AJ, 122, 163

\bibitem[{Gallagher {et~al.}(2010)Gallagher, Durrell, Elmegreen, Chandar,
  English, Charlton, Gronwall, Young, Tzanavaris, Johnson, de~Oliveira,
  Whitmore, Hornschemeier, Maybhate, , \& Zabludoff}]{gallagher10}
Gallagher, S.~C., Durrell, P.~R., Elmegreen, D.~M., Chandar, R., English, J.,
  Charlton, J.~C., Gronwall, C., Young, J., Tzanavaris, P., Johnson, K.~E.,
  de~Oliveira, C.~M., Whitmore, B., Hornschemeier, A.~E., Maybhate, A., , \&
  Zabludoff, A. 2010, The Astronomical Journal, 139, 545

\bibitem[{{Gallagher} {et~al.}(2008){Gallagher}, {Johnson}, {Hornschemeier},
  {Charlton}, \& {Hibbard}}]{gallagher08}
{Gallagher}, S.~C., {Johnson}, K.~E., {Hornschemeier}, A.~E., {Charlton},
  J.~C., \& {Hibbard}, J.~E. 2008, ApJ, 673, 730

\bibitem[{{Genzel} {et~al.}(2001){Genzel}, {Tacconi}, {Rigopoulou}, {Lutz}, \&
  {Tecza}}]{genzel01}
{Genzel}, R., {Tacconi}, L.~J., {Rigopoulou}, D., {Lutz}, D., \& {Tecza}, M.
  2001, ApJ, 563, 527

\bibitem[{{Goudfrooij} {et~al.}(2003){Goudfrooij}, {Strader}, {Brenneman},
  {Kissler-Patig}, {Minniti}, \& {Edwin Huizinga}}]{goudfrooij03}
{Goudfrooij}, P., {Strader}, J., {Brenneman}, L., {Kissler-Patig}, M.,
  {Minniti}, D., \& {Edwin Huizinga}, J. 2003, MNRAS, 343, 665

\bibitem[{{Harris}(1996)}]{harris96}
{Harris}, W.~E. 1996, AJ, 112, 1487

\bibitem[{{Harris}(2001)}]{harris01}
{Harris}, W.~E. 2001, in Saas-Fee Advanced Course 28: Star Clusters, ed.
  {L.~Labhardt \& B.~Binggeli}, 223--+

\bibitem[{{Harris} {et~al.}(2006){Harris}, {Whitmore}, {Karakla}, {Oko{\'n}},
  {Baum}, {Hanes}, \& {Kavelaars}}]{harris06}
{Harris}, W.~E., {Whitmore}, B.~C., {Karakla}, D., {Oko{\'n}}, W., {Baum},
  W.~A., {Hanes}, D.~A., \& {Kavelaars}, J.~J. 2006, ApJ, 636, 90

\bibitem[{{Haynes} \& {Giovanelli}(1984)}]{haynes84}
{Haynes}, M.~P., \& {Giovanelli}, R. 1984, AJ, 89, 758

\bibitem[{{Hibbard} \& {van Gorkom}(1996)}]{hibbard96}
{Hibbard}, J.~E., \& {van Gorkom}, J.~H. 1996, AJ, 111, 655

\bibitem[{{Hibbard} {et~al.}(2001){Hibbard}, {van Gorkom}, {Rupen}, \&
  {Schiminovich}}]{rogues}
{Hibbard}, J.~E., {van Gorkom}, J.~H., {Rupen}, M.~P., \& {Schiminovich}, D.
  2001, in Astronomical Society of the Pacific Conference Series, Vol. 240, Gas
  and Galaxy Evolution, ed. {J.~E.~Hibbard, M.~Rupen, \& J.~H.~van Gorkom},
  657--+

\bibitem[{{Hibbard} \& {Yun}(1999)}]{hibbard99}
{Hibbard}, J.~E., \& {Yun}, M.~S. 1999, AJ, 118, 162

\bibitem[{{Hickson}(1982)}]{hickson82}
{Hickson}, P. 1982, ApJ, 255, 382

\bibitem[{{Hickson} {et~al.}(1989){Hickson}, {Kindl}, \& {Auman}}]{hickson89}
{Hickson}, P., {Kindl}, E., \& {Auman}, J.~R. 1989, ApJS, 70, 687

\bibitem[{{Hickson} {et~al.}(1992){Hickson}, {Mendes de Oliveira}, {Huchra}, \&
  {Palumbo}}]{hickson92}
{Hickson}, P., {Mendes de Oliveira}, C., {Huchra}, J.~P., \& {Palumbo}, G.~G.
  1992, ApJ, 399, 353

\bibitem[{{Holtzman} {et~al.}(1992){Holtzman}, {Faber}, {Shaya}, {Lauer},
  {Groth}, {Hunter}, {Baum}, {Ewald}, {Hester}, {Light}, {Lynds}, {O'Neil}, \&
  {Westphal}}]{holtzman92}
{Holtzman}, J.~A., {Faber}, S.~M., {Shaya}, E.~J., {Lauer}, T.~R., {Groth}, J.,
  {Hunter}, D.~A., {Baum}, W.~A., {Ewald}, S.~P., {Hester}, J.~J., {Light},
  R.~M., {Lynds}, C.~R., {O'Neil}, Jr., E.~J., \& {Westphal}, J.~A. 1992, AJ,
  103, 691

\bibitem[{{Huchtmeier}(1997)}]{huchtmeier97}
{Huchtmeier}, W.~K. 1997, A\&A, 325, 473

\bibitem[{{Iglesias-P{\'a}ramo} \& {V{\'{\i}}lchez}(1999)}]{iglesias}
{Iglesias-P{\'a}ramo}, J., \& {V{\'{\i}}lchez}, J.~M. 1999, ApJ, 518, 94

\bibitem[{{Johnson} {et~al.}(2007){Johnson}, {Hibbard}, {Gallagher},
  {Charlton}, {Hornschemeier}, {Jarrett}, \& {Reines}}]{johnson07}
{Johnson}, K.~E., {Hibbard}, J.~E., {Gallagher}, S.~C., {Charlton}, J.~C.,
  {Hornschemeier}, A.~E., {Jarrett}, T.~H., \& {Reines}, A.~E. 2007, AJ, 134,
  1522

\bibitem[{{Kennicutt}(1998)}]{kennicutt98}
{Kennicutt}, Jr., R.~C. 1998, in ASPCS, Vol. 142, The Stellar Initial Mass
  Function (38th Herstmonceux Conference), ed. G.~{Gilmore} \& D.~{Howell},
  1--+

\bibitem[{{Kennicutt} {et~al.}(2003){Kennicutt}, {Armus}, {Bendo}, {Calzetti},
  {Dale}, {Draine}, {Engelbracht}, {Gordon}, {Grauer}, {Helou}, {Hollenbach},
  {Jarrett}, {Kewley}, {Leitherer}, {Li}, {Malhotra}, {Regan}, {Rieke},
  {Rieke}, {Roussel}, {Smith}, {Thornley}, \& {Walter}}]{sings}
{Kennicutt}, Jr., R.~C., {Armus}, L., {Bendo}, G., {Calzetti}, D., {Dale},
  D.~A., {Draine}, B.~T., {Engelbracht}, C.~W., {Gordon}, K.~D., {Grauer},
  A.~D., {Helou}, G., {Hollenbach}, D.~J., {Jarrett}, T.~H., {Kewley}, L.~J.,
  {Leitherer}, C., {Li}, A., {Malhotra}, S., {Regan}, M.~W., {Rieke}, G.~H.,
  {Rieke}, M.~J., {Roussel}, H., {Smith}, J., {Thornley}, M.~D., \& {Walter},
  F. 2003, PASP, 115, 928

\bibitem[{{Kennicutt} {et~al.}(2007){Kennicutt}, {Calzetti}, {Walter}, {Helou},
  {Hollenbach}, {Armus}, {Bendo}, {Dale}, {Draine}, {Engelbracht}, {Gordon},
  {Prescott}, {Regan}, {Thornley}, {Bot}, {Brinks}, {de Blok}, {de Mello},
  {Meyer}, {Moustakas}, {Murphy}, {Sheth}, \& {Smith}}]{kennicutt07}
{Kennicutt}, Jr., R.~C., {Calzetti}, D., {Walter}, F., {Helou}, G.,
  {Hollenbach}, D.~J., {Armus}, L., {Bendo}, G., {Dale}, D.~A., {Draine},
  B.~T., {Engelbracht}, C.~W., {Gordon}, K.~D., {Prescott}, M.~K.~M., {Regan},
  M.~W., {Thornley}, M.~D., {Bot}, C., {Brinks}, E., {de Blok}, E., {de Mello},
  D., {Meyer}, M., {Moustakas}, J., {Murphy}, E.~J., {Sheth}, K., \& {Smith},
  J.~D.~T. 2007, ApJ, 671, 333

\bibitem[{{Kewley} {et~al.}(2006){Kewley}, {Groves}, {Kauffmann}, \&
  {Heckman}}]{kewley06}
{Kewley}, L.~J., {Groves}, B., {Kauffmann}, G., \& {Heckman}, T. 2006, MNRAS,
  372, 961

\bibitem[{{Konstantopoulos}(2009)}]{isk09b}
{Konstantopoulos}, I.~S. 2009, ArXiv e-prints

\bibitem[{{Konstantopoulos} {et~al.}(2008){Konstantopoulos}, {Bastian},
  {Smith}, {Trancho}, {Westmoquette}, \& {Gallagher}}]{isk08}
{Konstantopoulos}, I.~S., {Bastian}, N., {Smith}, L.~J., {Trancho}, G.,
  {Westmoquette}, M.~S., \& {Gallagher}, III, J.~S. 2008, ApJ, 674, 846

\bibitem[{{Konstantopoulos} {et~al.}(2009){Konstantopoulos}, {Bastian},
  {Smith}, {Westmoquette}, {Trancho}, \& {Gallagher}}]{isk09a}
{Konstantopoulos}, I.~S., {Bastian}, N., {Smith}, L.~J., {Westmoquette}, M.~S.,
  {Trancho}, G., \& {Gallagher}, J.~S. 2009, ApJ, 701, 1015

\bibitem[{{Kormendy} \& {Kennicutt}(2004)}]{kk04}
{Kormendy}, J., \& {Kennicutt}, Jr., R.~C. 2004, ARA\&A, 42, 603

\bibitem[{{Krumholz} {et~al.}(2009){Krumholz}, {McKee}, \&
  {Tumlinson}}]{krumholz09}
{Krumholz}, M.~R., {McKee}, C.~F., \& {Tumlinson}, J. 2009, ApJL, 699, 850

\bibitem[{{Kundu} \& {Whitmore}(2001)}]{kundu01}
{Kundu}, A., \& {Whitmore}, B.~C. 2001, AJ, 122, 1251

\bibitem[{{Lada} \& {Lada}(2003)}]{ladalada03}
{Lada}, C.~J., \& {Lada}, E.~A. 2003, ARA\&A, 41, 57

\bibitem[{{Larsen}(2004)}]{larsen04}
{Larsen}, S.~S. 2004, A\&A, 416, 537

\bibitem[{{Leitherer} {et~al.}(1999){Leitherer}, {Schaerer}, {Goldader},
  {Gonz{\'a}lez Delgado}, {Robert}, {Kune}, {de Mello}, {Devost}, \&
  {Heckman}}]{sb99}
{Leitherer}, C., {Schaerer}, D., {Goldader}, J.~D., {Gonz{\'a}lez Delgado},
  R.~M., {Robert}, C., {Kune}, D.~F., {de Mello}, D.~F., {Devost}, D., \&
  {Heckman}, T.~M. 1999, ApJS, 123, 3

\bibitem[{{Leroy} {et~al.}(2008){Leroy}, {Walter}, {Brinks}, {Bigiel}, {de
  Blok}, {Madore}, \& {Thornley}}]{leroy08}
{Leroy}, A.~K., {Walter}, F., {Brinks}, E., {Bigiel}, F., {de Blok}, W.~J.~G.,
  {Madore}, B., \& {Thornley}, M.~D. 2008, AJ, 136, 2782

\bibitem[{{Lisenfeld} {et~al.}(2007){Lisenfeld}, {Verdes-Montenegro},
  {Sulentic}, {Leon}, {Espada}, {Bergond}, {Garc{\'{\i}}a}, {Sabater},
  {Santander-Vela}, \& {Verley}}]{amiga2}
{Lisenfeld}, U., {Verdes-Montenegro}, L., {Sulentic}, J., {Leon}, S., {Espada},
  D., {Bergond}, G., {Garc{\'{\i}}a}, E., {Sabater}, J., {Santander-Vela},
  J.~D., \& {Verley}, S. 2007, A\&A, 462, 507

\bibitem[{{Martig} \& {Bournaud}(2008)}]{martig08}
{Martig}, M., \& {Bournaud}, F. 2008, MNRAS, 385, L38

\bibitem[{{Mart{\'{\i}}nez} {et~al.}(2010){Mart{\'{\i}}nez}, {Del Olmo},
  {Coziol}, \& {Perea}}]{martinez10}
{Mart{\'{\i}}nez}, M.~A., {Del Olmo}, A., {Coziol}, R., \& {Perea}, J. 2010,
  AJ, 139, 1199

\bibitem[{{McConnachie} {et~al.}(2009){McConnachie}, {Irwin}, {Ibata},
  {Dubinski}, {Widrow}, {Martin}, {Cote}, {Dotter}, {Navarro}, {Ferguson},
  {Puzia}, {Lewis}, {Babul}, {Barmby}, {Bienayme}, {Chapman}, {Cockcroft},
  {Collins}, {Fardal}, {Harris}, {Huxor}, {Dougal Mackey}, {Penarrubia},
  {Rich}, {Richer}, {Siebert}, {Tanvir}, {Valls-Gabaud}, \&
  {Venn}}]{mcconnachie09}
{McConnachie}, A.~W., {Irwin}, M.~J., {Ibata}, R.~A., {Dubinski}, J., {Widrow},
  L.~M., {Martin}, N.~F., {Cote}, P., {Dotter}, A.~L., {Navarro}, J.~F.,
  {Ferguson}, A.~M.~N., {Puzia}, T.~H., {Lewis}, G.~F., {Babul}, A., {Barmby},
  P., {Bienayme}, O., {Chapman}, S.~C., {Cockcroft}, R., {Collins}, M.~L.~M.,
  {Fardal}, M.~A., {Harris}, W.~E., {Huxor}, A., {Dougal Mackey}, A.,
  {Penarrubia}, J., {Rich}, R.~M., {Richer}, H.~B., {Siebert}, A., {Tanvir},
  N., {Valls-Gabaud}, D., \& {Venn}, K.~A. 2009, ArXiv e-prints

\bibitem[{{McGaugh} {et~al.}(2000){McGaugh}, {Schombert}, {Bothun}, \& {de
  Blok}}]{mcgaugh00}
{McGaugh}, S.~S., {Schombert}, J.~M., {Bothun}, G.~D., \& {de Blok}, W.~J.~G.
  2000, ApJL, 533, L99

\bibitem[{{Memola} {et~al.}(2009){Memola}, {Trinchieri}, {Wolter}, {Focardi},
  \& {Kelm}}]{memola09}
{Memola}, E., {Trinchieri}, G., {Wolter}, A., {Focardi}, P., \& {Kelm}, B.
  2009, A\&A, 497, 359

\bibitem[{{Mendes de Oliveira} \& {Hickson}(1994)}]{mendes94}
{Mendes de Oliveira}, C., \& {Hickson}, P. 1994, ApJ, 427, 684

\bibitem[{{Mihos}(2001)}]{mihos01}
{Mihos}, J.~C. 2001, ApJ, 550, 94

\bibitem[{{Moles} {et~al.}(1994){Moles}, {del Olmo}, {Perea}, {Masegosa},
  {Marquez}, \& {Costa}}]{moles94}
{Moles}, M., {del Olmo}, A., {Perea}, J., {Masegosa}, J., {Marquez}, I., \&
  {Costa}, V. 1994, A\&A, 285, 404

\bibitem[{{Muratov} \& {Gnedin}(2010)}]{muratov10}
{Muratov}, A.~L., \& {Gnedin}, O.~Y. 2010, ArXiv e-prints

\bibitem[{{Oestlin} {et~al.}(1998){Oestlin}, {Bergvall}, \&
  {Roennback}}]{oestlin98}
{Oestlin}, G., {Bergvall}, N., \& {Roennback}, J. 1998, A\&A, 335, 85

\bibitem[{{{\"O}stlin} {et~al.}(2003){{\"O}stlin}, {Zackrisson}, {Bergvall}, \&
  {R{\"o}nnback}}]{oestlin03}
{{\"O}stlin}, G., {Zackrisson}, E., {Bergvall}, N., \& {R{\"o}nnback}, J. 2003,
  A\&A, 408, 887

\bibitem[{{O'Sullivan} {et~al.}(2007){O'Sullivan}, {Sanderson}, \&
  {Ponman}}]{osullivan07}
{O'Sullivan}, E., {Sanderson}, A.~J.~R., \& {Ponman}, T.~J. 2007, MNRAS, 380,
  1409

\bibitem[{{Palma} {et~al.}(2002){Palma}, {Zonak}, {Hunsberger}, {Charlton},
  {Gallagher}, {Durrell}, \& {English}}]{palma}
{Palma}, C., {Zonak}, S.~G., {Hunsberger}, S.~D., {Charlton}, J.~C.,
  {Gallagher}, S.~C., {Durrell}, P.~R., \& {English}, J. 2002, AJ, 124, 2425

\bibitem[{{Peng} {et~al.}(2006){Peng}, {Jord{\'a}n}, {C{\^o}t{\'e}},
  {Blakeslee}, {Ferrarese}, {Mei}, {West}, {Merritt}, {Milosavljevi{\'c}}, \&
  {Tonry}}]{peng06}
{Peng}, E.~W., {Jord{\'a}n}, A., {C{\^o}t{\'e}}, P., {Blakeslee}, J.~P.,
  {Ferrarese}, L., {Mei}, S., {West}, M.~J., {Merritt}, D.,
  {Milosavljevi{\'c}}, M., \& {Tonry}, J.~L. 2006, ApJ, 639, 95

\bibitem[{{Plana} {et~al.}(1998){Plana}, {Mendes de Oliveira}, {Amram}, \&
  {Boulesteix}}]{plana98}
{Plana}, H., {Mendes de Oliveira}, C., {Amram}, P., \& {Boulesteix}, J. 1998,
  AJ, 116, 2123

\bibitem[{{Rasmussen} {et~al.}(2008){Rasmussen}, {Ponman}, {Verdes-Montenegro},
  {Yun}, \& {Borthakur}}]{rasmussen08}
{Rasmussen}, J., {Ponman}, T.~J., {Verdes-Montenegro}, L., {Yun}, M.~S., \&
  {Borthakur}, S. 2008, MNRAS, 388, 1245

\bibitem[{{Rejkuba} {et~al.}(2005){Rejkuba}, {Greggio}, {Harris}, {Harris}, \&
  {Peng}}]{rejkuba05}
{Rejkuba}, M., {Greggio}, L., {Harris}, W.~E., {Harris}, G.~L.~H., \& {Peng},
  E.~W. 2005, ApJ, 631, 262

\bibitem[{{Rhode} {et~al.}(2007){Rhode}, {Zepf}, {Kundu}, \&
  {Larner}}]{rhode07}
{Rhode}, K.~L., {Zepf}, S.~E., {Kundu}, A., \& {Larner}, A.~N. 2007, AJ, 134,
  1403

\bibitem[{{Robin} {et~al.}(2003){Robin}, {Reyl{\'e}}, {Derri{\`e}re}, \&
  {Picaud}}]{robin03}
{Robin}, A.~C., {Reyl{\'e}}, C., {Derri{\`e}re}, S., \& {Picaud}, S. 2003,
  A\&A, 409, 523

\bibitem[{{Roming} {et~al.}(2005){Roming}, {Kennedy}, {Mason}, {Nousek}, {Ahr},
  {Bingham}, {Broos}, {Carter}, {Hancock}, {Huckle}, {Hunsberger}, {Kawakami},
  {Killough}, {Koch}, {McLelland}, {Smith}, {Smith}, {Soto}, {Boyd},
  {Breeveld}, {Holland}, {Ivanushkina}, {Pryzby}, {Still}, \&
  {Stock}}]{roming05}
{Roming}, P.~W.~A., {Kennedy}, T.~E., {Mason}, K.~O., {Nousek}, J.~A., {Ahr},
  L., {Bingham}, R.~E., {Broos}, P.~S., {Carter}, M.~J., {Hancock}, B.~K.,
  {Huckle}, H.~E., {Hunsberger}, S.~D., {Kawakami}, H., {Killough}, R., {Koch},
  T.~S., {McLelland}, M.~K., {Smith}, K., {Smith}, P.~J., {Soto}, J.~C.,
  {Boyd}, P.~T., {Breeveld}, A.~A., {Holland}, S.~T., {Ivanushkina}, M.,
  {Pryzby}, M.~S., {Still}, M.~D., \& {Stock}, J. 2005, Space Science Reviews,
  120, 95

\bibitem[{{Rubin} {et~al.}(1991){Rubin}, {Hunter}, \& {Ford}}]{rubin91}
{Rubin}, V.~C., {Hunter}, D.~A., \& {Ford}, W.~K.~J. 1991, ApJS, 76, 153

\bibitem[{{Salzer} {et~al.}(2005){Salzer}, {Jangren}, {Gronwall}, {Werk},
  {Chomiuk}, {Caperton}, {Melbourne}, \& {McKinstry}}]{salzer05}
{Salzer}, J.~J., {Jangren}, A., {Gronwall}, C., {Werk}, J.~K., {Chomiuk},
  L.~B., {Caperton}, K.~A., {Melbourne}, J., \& {McKinstry}, K. 2005, AJ, 130,
  2584

\bibitem[{{Sanders} \& {Mirabel}(1996)}]{sanders96}
{Sanders}, D.~B., \& {Mirabel}, I.~F. 1996, ARA\&A, 34, 749

\bibitem[{{Scheepmaker} {et~al.}(2007){Scheepmaker}, {Haas}, {Gieles},
  {Bastian}, {Larsen}, \& {Lamers}}]{remco07}
{Scheepmaker}, R.~A., {Haas}, M.~R., {Gieles}, M., {Bastian}, N., {Larsen},
  S.~S., \& {Lamers}, H.~J.~G.~L.~M. 2007, A\&A, 469, 925

\bibitem[{{Schlegel} {et~al.}(1998){Schlegel}, {Finkbeiner}, \&
  {Davis}}]{schlegel98}
{Schlegel}, D.~J., {Finkbeiner}, D.~P., \& {Davis}, M. 1998, ApJ, 500, 525

\bibitem[{{Schweizer} \& {Seitzer}(1998)}]{schweizer98}
{Schweizer}, F., \& {Seitzer}, P. 1998, AJ, 116, 2206

\bibitem[{{Silva} {et~al.}(1998){Silva}, {Granato}, {Bressan}, \&
  {Danese}}]{silva98}
{Silva}, L., {Granato}, G.~L., {Bressan}, A., \& {Danese}, L. 1998, ApJ, 509,
  103

\bibitem[{{Sirianni} {et~al.}(2005){Sirianni}, {Jee}, {Ben{\'{\i}}tez},
  {Blakeslee}, {Martel}, {Meurer}, {Clampin}, {De Marchi}, {Ford}, {Gilliland},
  {Hartig}, {Illingworth}, {Mack}, \& {McCann}}]{sirianni05}
{Sirianni}, M., {Jee}, M.~J., {Ben{\'{\i}}tez}, N., {Blakeslee}, J.~P.,
  {Martel}, A.~R., {Meurer}, G., {Clampin}, M., {De Marchi}, G., {Ford}, H.~C.,
  {Gilliland}, R., {Hartig}, G.~F., {Illingworth}, G.~D., {Mack}, J., \&
  {McCann}, W.~J. 2005, PASP, 117, 1049

\bibitem[{{Skrutskie} {et~al.}(2006){Skrutskie}, {Cutri}, {Stiening},
  {Weinberg}, {Schneider}, {Carpenter}, {Beichman}, {Capps}, {Chester},
  {Elias}, {Huchra}, {Liebert}, {Lonsdale}, {Monet}, {Price}, {Seitzer},
  {Jarrett}, {Kirkpatrick}, {Gizis}, {Howard}, {Evans}, {Fowler}, {Fullmer},
  {Hurt}, {Light}, {Kopan}, {Marsh}, {McCallon}, {Tam}, {Van Dyk}, \&
  {Wheelock}}]{2mass}
{Skrutskie}, M.~F., {Cutri}, R.~M., {Stiening}, R., {Weinberg}, M.~D.,
  {Schneider}, S., {Carpenter}, J.~M., {Beichman}, C., {Capps}, R., {Chester},
  T., {Elias}, J., {Huchra}, J., {Liebert}, J., {Lonsdale}, C., {Monet}, D.~G.,
  {Price}, S., {Seitzer}, P., {Jarrett}, T., {Kirkpatrick}, J.~D., {Gizis},
  J.~E., {Howard}, E., {Evans}, T., {Fowler}, J., {Fullmer}, L., {Hurt}, R.,
  {Light}, R., {Kopan}, E.~L., {Marsh}, K.~A., {McCallon}, H.~L., {Tam}, R.,
  {Van Dyk}, S., \& {Wheelock}, S. 2006, AJ, 131, 1163

\bibitem[{{Solomon} {et~al.}(1987){Solomon}, {Rivolo}, {Barrett}, \&
  {Yahil}}]{solomon87}
{Solomon}, P.~M., {Rivolo}, A.~R., {Barrett}, J., \& {Yahil}, A. 1987, ApJ,
  319, 730

\bibitem[{{Spergel} {et~al.}(2007){Spergel}, {Bean}, {Dor{\'e}}, {Nolta},
  {Bennett}, {Dunkley}, {Hinshaw}, {Jarosik}, {Komatsu}, {Page}, {Peiris},
  {Verde}, {Halpern}, {Hill}, {Kogut}, {Limon}, {Meyer}, {Odegard}, {Tucker},
  {Weiland}, {Wollack}, \& {Wright}}]{spergel07}
{Spergel}, D.~N., {Bean}, R., {Dor{\'e}}, O., {Nolta}, M.~R., {Bennett}, C.~L.,
  {Dunkley}, J., {Hinshaw}, G., {Jarosik}, N., {Komatsu}, E., {Page}, L.,
  {Peiris}, H.~V., {Verde}, L., {Halpern}, M., {Hill}, R.~S., {Kogut}, A.,
  {Limon}, M., {Meyer}, S.~S., {Odegard}, N., {Tucker}, G.~S., {Weiland},
  J.~L., {Wollack}, E., \& {Wright}, E.~L. 2007, \apjs, 170, 377

\bibitem[{{Tago} {et~al.}(2008){Tago}, {Einasto}, {Saar}, {Tempel}, {Einasto},
  {Vennik}, \& {M{\"u}ller}}]{tago08}
{Tago}, E., {Einasto}, J., {Saar}, E., {Tempel}, E., {Einasto}, M., {Vennik},
  J., \& {M{\"u}ller}, V. 2008, A\&A, 479, 927

\bibitem[{{The} \& {White}(1986)}]{the86}
{The}, L.~S., \& {White}, S.~D.~M. 1986, AJ, 92, 1248

\bibitem[{Torres-Flores {et~al.}(2009)Torres-Flores, Mendes~de Oliveira,
  de~Mello, Amram, Plana, Epinat, \& Iglesias-P\'aramo}]{torres09}
Torres-Flores, S., Mendes~de Oliveira, C., de~Mello, D.~F., Amram, P., Plana,
  H., Epinat, B., \& Iglesias-P\'aramo, J. 2009, A\&A, 507, 723

\bibitem[{{Trancho} {et~al.}(2007{\natexlab{a}}){Trancho}, {Bastian}, {Miller},
  \& {Schweizer}}]{gelys07b}
{Trancho}, G., {Bastian}, N., {Miller}, B.~W., \& {Schweizer}, F.
  2007{\natexlab{a}}, ApJ, 664, 284

\bibitem[{{Trancho} {et~al.}(2007{\natexlab{b}}){Trancho}, {Bastian},
  {Schweizer}, \& {Miller}}]{gelys07a}
{Trancho}, G., {Bastian}, N., {Schweizer}, F., \& {Miller}, B.~W.
  2007{\natexlab{b}}, ApJ, 658, 993

\bibitem[{{Tzanavaris} {et~al.}(2010){Tzanavaris}, {Hornschemeier},
  {Gallagher}, {Johnson}, {Gronwall}, {Immler}, {Reines}, {Hoversten}, \&
  {Charlton}}]{tzanavaris10}
{Tzanavaris}, P., {Hornschemeier}, A.~E., {Gallagher}, S.~C., {Johnson}, K.~E.,
  {Gronwall}, C., {Immler}, S., {Reines}, A.~E., {Hoversten}, E., \&
  {Charlton}, J.~C. 2010, ApJ, 716, 556

\bibitem[{{van Dokkum}(2001)}]{lacosmic}
{van Dokkum}, P.~G. 2001, PASP, 113, 1420

\bibitem[{{van Dokkum}(2005)}]{vandokkum05}
---. 2005, AJ, 130, 2647

\bibitem[{{Verdes-Montenegro} {et~al.}(2005){Verdes-Montenegro}, {Sulentic},
  {Lisenfeld}, {Leon}, {Espada}, {Garcia}, {Sabater}, \& {Verley}}]{amiga1}
{Verdes-Montenegro}, L., {Sulentic}, J., {Lisenfeld}, U., {Leon}, S., {Espada},
  D., {Garcia}, E., {Sabater}, J., \& {Verley}, S. 2005, A\&A, 436, 443

\bibitem[{{Verdes-Montenegro} {et~al.}(1998){Verdes-Montenegro}, {Yun},
  {Perea}, {del Olmo}, \& {Ho}}]{vm98}
{Verdes-Montenegro}, L., {Yun}, M.~S., {Perea}, J., {del Olmo}, A., \& {Ho},
  P.~T.~P. 1998, ApJ, 497, 89

\bibitem[{{Verdes-Montenegro} {et~al.}(2001){Verdes-Montenegro}, {Yun},
  {Williams}, {Huchtmeier}, {Del Olmo}, \& {Perea}}]{verdes01}
{Verdes-Montenegro}, L., {Yun}, M.~S., {Williams}, B.~A., {Huchtmeier}, W.~K.,
  {Del Olmo}, A., \& {Perea}, J. 2001, A\&A, 377, 812

\bibitem[{{Walker} {et~al.}(2009){Walker}, {Johnson}, {Gallagher}, {Hibbard},
  {Hornschemeier}, {Charlton}, \& {Jarrett}}]{walker09}
{Walker}, L.~M., {Johnson}, K.~E., {Gallagher}, S.~C., {Hibbard}, J.~E.,
  {Hornschemeier}, A.~E., {Charlton}, J.~C., \& {Jarrett}, T.~H. 2009, ArXiv
  e-prints

\bibitem[{{Walker} {et~al.}(2010){Walker}, {Johnson}, {Gallagher}, {Hibbard},
  {Hornschemeier}, {Charlton}, \& {Jarrett}}]{walker10}
{Walker}, L.~M., {Johnson}, K.~E., {Gallagher}, S.~C., {Hibbard}, J.~E.,
  {Hornschemeier}, A.~E., {Charlton}, J.~C., \& {Jarrett}, T.~H. 2010, in
  Bulletin of the American Astronomical Society, Vol.~41, Bulletin of the
  American Astronomical Society, 237--+

\bibitem[{{White} {et~al.}(2003){White}, {Bothun}, {Guerrero}, {West}, \&
  {Barkhouse}}]{white}
{White}, P.~M., {Bothun}, G., {Guerrero}, M.~A., {West}, M.~J., \& {Barkhouse},
  W.~A. 2003, ApJ, 585, 739

\bibitem[{{Whitmore} {et~al.}(2010){Whitmore}, {Chandar}, {Schweizer},
  {Rothberg}, {Leitherer}, {Rieke}, {Rieke}, {Blair}, {Mengel}, \&
  {Alonso-Herrero}}]{whitmore10}
{Whitmore}, B.~C., {Chandar}, R., {Schweizer}, F., {Rothberg}, B., {Leitherer},
  C., {Rieke}, M., {Rieke}, G., {Blair}, W.~P., {Mengel}, S., \&
  {Alonso-Herrero}, A. 2010, ArXiv e-prints

\bibitem[{{Whitmore} {et~al.}(1999){Whitmore}, {Zhang}, {Leitherer}, {Fall},
  {Schweizer}, \& {Miller}}]{whitmore99}
{Whitmore}, B.~C., {Zhang}, Q., {Leitherer}, C., {Fall}, S.~M., {Schweizer},
  F., \& {Miller}, B.~W. 1999, AJ, 118, 1551

\bibitem[{{Williams} \& {Rood}(1987)}]{williams87}
{Williams}, B.~A., \& {Rood}, H.~J. 1987, ApJS, 63, 265

\bibitem[{{Wilman} {et~al.}(2009){Wilman}, {Oemler}, {Mulchaey}, {McGee},
  {Balogh}, \& {Bower}}]{wilman}
{Wilman}, D.~J., {Oemler}, A., {Mulchaey}, J.~S., {McGee}, S.~L., {Balogh},
  M.~L., \& {Bower}, R.~G. 2009, ApJ, 692, 298

\bibitem[{{York} {et~al.}(2000){York}, {Adelman}, {Anderson}, {Anderson},
  {Annis}, {Bahcall}, {Bakken}, {Barkhouser}, {Bastian}, {Berman}, {Boroski},
  {Bracker}, {Briegel}, {Briggs}, {Brinkmann}, {Brunner}, {Burles}, {Carey},
  {Carr}, {Castander}, {Chen}, {Colestock}, {Connolly}, {Crocker}, {Csabai},
  {Czarapata}, {Davis}, {Doi}, {Dombeck}, {Eisenstein}, {Ellman}, {Elms},
  {Evans}, {Fan}, {Federwitz}, {Fiscelli}, {Friedman}, {Frieman}, {Fukugita},
  {Gillespie}, {Gunn}, {Gurbani}, {de Haas}, {Haldeman}, {Harris}, {Hayes},
  {Heckman}, {Hennessy}, {Hindsley}, {Holm}, {Holmgren}, {Huang}, {Hull},
  {Husby}, {Ichikawa}, {Ichikawa}, {Ivezi{\'c}}, {Kent}, {Kim}, {Kinney},
  {Klaene}, {Kleinman}, {Kleinman}, {Knapp}, {Korienek}, {Kron}, {Kunszt},
  {Lamb}, {Lee}, {Leger}, {Limmongkol}, {Lindenmeyer}, {Long}, {Loomis},
  {Loveday}, {Lucinio}, {Lupton}, {MacKinnon}, {Mannery}, {Mantsch}, {Margon},
  {McGehee}, {McKay}, {Meiksin}, {Merelli}, {Monet}, {Munn}, {Narayanan},
  {Nash}, {Neilsen}, {Neswold}, {Newberg}, {Nichol}, {Nicinski}, {Nonino},
  {Okada}, {Okamura}, {Ostriker}, {Owen}, {Pauls}, {Peoples}, {Peterson},
  {Petravick}, {Pier}, {Pope}, {Pordes}, {Prosapio}, {Rechenmacher}, {Quinn},
  {Richards}, {Richmond}, {Rivetta}, {Rockosi}, {Ruthmansdorfer}, {Sandford},
  {Schlegel}, {Schneider}, {Sekiguchi}, {Sergey}, {Shimasaku}, {Siegmund},
  {Smee}, {Smith}, {Snedden}, {Stone}, {Stoughton}, {Strauss}, {Stubbs},
  {SubbaRao}, {Szalay}, {Szapudi}, {Szokoly}, {Thakar}, {Tremonti}, {Tucker},
  {Uomoto}, {Vanden Berk}, {Vogeley}, {Waddell}, {Wang}, {Watanabe},
  {Weinberg}, {Yanny}, \& {Yasuda}}]{sdss}
{York}, D.~G., {Adelman}, J., {Anderson}, Jr., J.~E., {Anderson}, S.~F.,
  {Annis}, J., {Bahcall}, N.~A., {Bakken}, J.~A., {Barkhouser}, R., {Bastian},
  S., {Berman}, E., {Boroski}, W.~N., {Bracker}, S., {Briegel}, C., {Briggs},
  J.~W., {Brinkmann}, J., {Brunner}, R., {Burles}, S., {Carey}, L., {Carr},
  M.~A., {Castander}, F.~J., {Chen}, B., {Colestock}, P.~L., {Connolly}, A.~J.,
  {Crocker}, J.~H., {Csabai}, I., {Czarapata}, P.~C., {Davis}, J.~E., {Doi},
  M., {Dombeck}, T., {Eisenstein}, D., {Ellman}, N., {Elms}, B.~R., {Evans},
  M.~L., {Fan}, X., {Federwitz}, G.~R., {Fiscelli}, L., {Friedman}, S.,
  {Frieman}, J.~A., {Fukugita}, M., {Gillespie}, B., {Gunn}, J.~E., {Gurbani},
  V.~K., {de Haas}, E., {Haldeman}, M., {Harris}, F.~H., {Hayes}, J.,
  {Heckman}, T.~M., {Hennessy}, G.~S., {Hindsley}, R.~B., {Holm}, S.,
  {Holmgren}, D.~J., {Huang}, C., {Hull}, C., {Husby}, D., {Ichikawa}, S.,
  {Ichikawa}, T., {Ivezi{\'c}}, {\v Z}., {Kent}, S., {Kim}, R.~S.~J., {Kinney},
  E., {Klaene}, M., {Kleinman}, A.~N., {Kleinman}, S., {Knapp}, G.~R.,
  {Korienek}, J., {Kron}, R.~G., {Kunszt}, P.~Z., {Lamb}, D.~Q., {Lee}, B.,
  {Leger}, R.~F., {Limmongkol}, S., {Lindenmeyer}, C., {Long}, D.~C., {Loomis},
  C., {Loveday}, J., {Lucinio}, R., {Lupton}, R.~H., {MacKinnon}, B.,
  {Mannery}, E.~J., {Mantsch}, P.~M., {Margon}, B., {McGehee}, P., {McKay},
  T.~A., {Meiksin}, A., {Merelli}, A., {Monet}, D.~G., {Munn}, J.~A.,
  {Narayanan}, V.~K., {Nash}, T., {Neilsen}, E., {Neswold}, R., {Newberg},
  H.~J., {Nichol}, R.~C., {Nicinski}, T., {Nonino}, M., {Okada}, N., {Okamura},
  S., {Ostriker}, J.~P., {Owen}, R., {Pauls}, A.~G., {Peoples}, J., {Peterson},
  R.~L., {Petravick}, D., {Pier}, J.~R., {Pope}, A., {Pordes}, R., {Prosapio},
  A., {Rechenmacher}, R., {Quinn}, T.~R., {Richards}, G.~T., {Richmond}, M.~W.,
  {Rivetta}, C.~H., {Rockosi}, C.~M., {Ruthmansdorfer}, K., {Sandford}, D.,
  {Schlegel}, D.~J., {Schneider}, D.~P., {Sekiguchi}, M., {Sergey}, G.,
  {Shimasaku}, K., {Siegmund}, W.~A., {Smee}, S., {Smith}, J.~A., {Snedden},
  S., {Stone}, R., {Stoughton}, C., {Strauss}, M.~A., {Stubbs}, C., {SubbaRao},
  M., {Szalay}, A.~S., {Szapudi}, I., {Szokoly}, G.~P., {Thakar}, A.~R.,
  {Tremonti}, C., {Tucker}, D.~L., {Uomoto}, A., {Vanden Berk}, D., {Vogeley},
  M.~S., {Waddell}, P., {Wang}, S., {Watanabe}, M., {Weinberg}, D.~H., {Yanny},
  B., \& {Yasuda}, N. 2000, AJ, 120, 1579

\bibitem[{{Zabludoff} \& {Mulchaey}(1998)}]{zab98}
{Zabludoff}, A.~I., \& {Mulchaey}, J.~S. 1998, ApJ, 496, 39

\bibitem[{{Zabludoff} {et~al.}(1996){Zabludoff}, {Zaritsky}, {Lin}, {Tucker},
  {Hashimoto}, {Shectman}, {Oemler}, \& {Kirshner}}]{zabludoff96}
{Zabludoff}, A.~I., {Zaritsky}, D., {Lin}, H., {Tucker}, D., {Hashimoto}, Y.,
  {Shectman}, S.~A., {Oemler}, A., \& {Kirshner}, R.~P. 1996, ApJ, 466, 104

\bibitem[{{Zhang} \& {Fall}(1999)}]{zhang99}
{Zhang}, Q., \& {Fall}, S.~M. 1999, ApJL, 527, L81

\end{thebibliography}
\clearpage

\end{document}